\documentclass[journal,10pt]{IEEEtran}

\usepackage[usenames,dvipsnames]{color}
\usepackage{algorithm,algorithmic}
\usepackage[cmex10]{amsmath}
\usepackage[pdftex]{graphicx}
\usepackage{verbatim}
\usepackage{amsfonts}
\usepackage{soul}
\usepackage{rotating}

\usepackage{booktabs}
\usepackage{color}

\usepackage{graphics}
\usepackage{subfigure}
\ifCLASSINFOpdf
   \graphicspath{{}}
   \DeclareGraphicsExtensions{.pdf}
\else

\fi

\normalsize

\newfont{\bb}{msbm10 scaled 1100}
\newcommand{\vet}[1]{\mathbf{#1}}

\newcommand{\mol}[1]{\ensuremath{\textrm{#1}}}

\newcommand{\sect}{Sect.}
\newcommand{\fig}{Fig.~}

\newcommand{\stoicMat}{\ensuremath{\mathbf{\Xi}}}

\newcommand{\algAC}{\ensuremath{\mathcal{A}}}
\newcommand{\AC}{\ensuremath{\mathcal{AC}}}

\newcommand{\conc}[1]{\ensuremath{c_{\textrm{#1}}}}

\def\etc{\emph{etc.}}

\def\lista{\emph{(i)}~}
\def\listb{\emph{(ii)}~}
\def\listc{\emph{(iii)}~}
\def\listd{\emph{(iv)}~}
\def\liste{\emph{(v)}~}

\def\cmem{{$c$}-\texttt{mem} }
\def\amem{{$\alpha$}-\texttt{mem} }
\def\bmem{{$\beta$}-\texttt{mem} }
\def\kmem{{$k$}-\texttt{mem} }

\newcommand{\newtext}[1]{{\color{black}#1}}

\begin{document}

\title{\huge{Towards Programmable Network Dynamics:\\A Chemistry-Inspired Abstraction for Hardware Design}}

\author{Massimo~Monti, \emph{Student Member IEEE}, 
Manolis Sifalakis, \emph{Member IEEE},
Christian~F.~Tschudin, \emph{Member IEEE},
and Marco~Luise, \emph{Fellow IEEE}

\thanks{M.~Monti, M.~Sifalakis, and C.~Tschudin are with the Department of
Mathematics and Computer Science, University of Basel, Bernoullistrasse 16,
4056 Basel, Switzerland
(e-mail:\{m.monti,sifalakis.manos,christian.tschudin\}@unibas.ch).}

\thanks{M.~Luise is with the Information Engineering Department, University of
Pisa, Via Caruso 16, 56126 Pisa, Italy (e-mail:marco.luise@iet.unipi.it).} 

\thanks{This work has been supported in part by Swiss National Science
Foundation grant \#132525.}

} \maketitle

\begin{abstract} 

Chemical algorithms are statistical algorithms described and represented as
chemical reaction networks. They are particularly attractive for traffic
shaping and general control of network dynamics; they are analytically
tractable, they reinforce a strict state-to-dynamics relationship, they have
configurable stability properties, and they are directly implemented in
state-space using a high-level (graphical) representation.
In this paper, we present a direct implementation of chemical algorithms on
FPGA hardware. Besides substantially improving performance, we have achieved
hardware-level programmability and re-configurability of these algorithms
\emph{at runtime} (not interrupting servicing) and \emph{in realtime} (with
sub-second latency). This opens an interesting perspective for expanding the
currently limited scope of software defined networking and network
virtualisation solutions, to include programmable control of network dynamics.

\end{abstract}


\begin{IEEEkeywords}
Chemical algorithm, 
Programmable networks,
Software defined networking, 
Network dynamics, 
Traffic shaping, 
FPGA.
\end{IEEEkeywords}



\section{Introduction}
\label{sec:intro}

\IEEEPARstart{N}{etwork dynamics}
\newtext{as a term describes an important operational aspect of queuing
networks and the Internet. It refers to traffic control processes such as
(among others) scheduling, shaping, policing, and Active Queue Management
(AQM).} 
Initially, network dynamics were controlled end-to-end only, through transport
protocol mechanisms such as TCP's flow and congestion control algorithms. Yet
today, also mechanisms at the core of the network play an important role in
shaping intra/inter-flow dynamics in the Internet. This is effected for example by
means of service differentiation, flow conditioning, (distributed) rate control,
AQM and other congestion avoidance measures.

\newtext{
Recent developments in network virtualisation for cloud infrastructures and
Software Defined Networking (SDN), which explore ways to make network
infrastructure runtime-volatile through software, have at large neglected
network dynamics. To date, most of efforts have focused on
programmability of data paths (functions pertinent to
firewalling, packet inspection, header editing, \etc) and topology management.
In regard to network dynamics on the other hand, while notable advancements
are being underway today
\cite{AlKa12}\nocite{AlGr10,RoSa11,JeAl13,BaCo11,NiJa12,PaNa12}--\cite{ThBa13},
programmable SDN-like deployments are at best confined to a number of
\emph{pre-packaged} (often ``age-old'') algorithms, typically offered as
proprietary manufacturer-provided modules \cite{SDNet}. A
potentially notable exception is the software-switch specification language proposed
in~\cite{BoDa14}.
Authors claim that it may be used to create action primitives for congestion control 
although they fall short of explaining how (or providing examples).
}

Enabling runtime programmability/configurability of functions to control
network dynamics is more challenging
%
than accessing and modifying the router/node fabric to simply extend
packet parsing and filtering functionalities (e.g., \cite{JeAl14,BoGi13}) or
perform topology management (e.g., \cite{EthSwicth}).
%
First of all, it requires solutions that can be deployed close to, or on,
hardware (for performance and computational speed reasons). Additionally, in contrast to a mere flow-rule pipeline, such
functions are algorithmically complex to implement, with many interdependencies
to cater for. For example, programming
or reconfiguring a queueing discipline often requires to modify the actual
logic \cite{NiJa12,PaNa12,Ku13} that functionally binds
different runtime parameters and components (e.g.  queue-lengths, filter
thresholds, droppers and markers, averaging coefficients, \etc). Next,
management operations for modifying parameters in these functions (e.g.
setting rate cap parameters, meter bands, \etc) are likely more frequent than
typical topology management tasks. And finally, changes (not only modifications
but also the replacement of algorithms) are less tolerant to data-path delays than load
operations of flow-table rules.


In past works~\cite{hotnet09}--\nocite{sensorj14, srds12, jsac13,danms13}\cite{ijcnc13}, we have introduced a class of algorithms founded on
operational principles of chemical reaction networks, and demonstrated their
suitability for (expressible state-space representation) and usefulness in
(analysability/verification) the design of control functions for various
tasks pertinent to network dynamics.
In this paper, we capitalise on, and complete, this work in the context of
programmable networks, and we show that these ``chemical'' algorithms (CAs) are
fast to deploy and easy to re-program and modify \emph{at runtime} on FPGA
hardware.
\newtext{
Specifically the contributions of this work amount to the following:

\begin{enumerate}
\item Direct expressibility of high level mathematical models of control
systems on hardware, based on the simple reaction network
abstraction, without resorting to cumbersome hardware
description language (HDL) programming.

\item Effective algorithmic parallelisation without special engineering effort
or the need for compiler optimisation. As these models freely describe
parallelisable logic in state equations, they do not need to be implemented as
finite state automata (i.e.  sequentialised algorithms) so that they can
be executed by a CPU (soft- or hard-). 

\item Re-programmability (parameter tuning, partial algorithm rewriting, but
also complete algorithm replacement) on hardware, \emph{at runtime} and
\emph{sub-second latencies}, without need for bitstream re-generation and
re-loading on the FPGA. In principle (albeit not experimented) our method
should enable re-programmability of such algorithms even on ASICs.
\end{enumerate}

The implementation of CAs on hardware opens an unprecedented possibility in
SDN and programmable networks to support customisable
network dynamics functions, with fast prototyping, fast deployment, prompt
testing and verification. To our knowledge, we have not seen a record of
similar or analogous contribution in this field so far.
}

\newtext{
The rest of this paper is organised as follows. In the next subsection we
motivate our work {and clarify our contributions}.  In
section \sect\ref{sec:AC} we \newtext{summarise} the basics of CAs to a certain degree of
detail as essential to explain our design on hardware. In
\sect\ref{sec:FPGAimplHW} we present our framework design for running CAs on
FPGA hardware and in \sect\ref{sec:expHW} we evaluate an implementation on the
Xilinx Spartan-6 XC6SLX9 FPGA device. {Finally in
\sect\ref{sec:discussion} we give examples of CAs for queue management (not previously presented in the
literature), we discuss in the context of SDN an integration approach for our
framework in the OpenFlow architecture~\cite{McPa08}, and we provide an account
of what performance can one anticipate with the FPGA technology currently available
on market.}
}


\subsection{Motivation}
\label{sec:motivation}

\newtext{
Research and engineering efforts in SDN and virtualisation for cloud
infrastructures have been exploring ways to make network infrastructure
run-time volatile through software. The aim is to simplify network management
and improve service provisioning in response to fast-changing user demand,
mobility, distributed multipoint access, \etc. So far, most of research
in SDN has focused on defining open protocols and interfaces to create a
very generic and flexible switch architecture
(capable of accommodating bespoke packet processing functions). In this process, there have often been attempts to bring ideas and solutions from active/programmable networks closer to (FPGA)
hardware, so to address concerns on delay performance and processing off-loading (from CPU).
Classic examples of projects in this direction have been the NetFPGA
large-scale collaborative initiative~\cite{LoMc07} (among universities and FPGA
manufacturers) as well as works taking place in individual labs of IC
manufacturers~\cite{SDNet, EthSwicth}, which develop \emph{Intellectual Property
Cores} (IPCs) for complex networking functions that can be used off-the-shelf
in the synthesis of composable data-planes on FPGAs.

In the whole volume of work that exists so far, we are able to identify two
important issues. One is the lack of (or limited) attention to the
programmability of the parts of the inter-networking fabric that pertains to
traffic management and network dynamics in general (we have only seen the topic
touched in~\cite{SDNet, BoDa14,
AlteraOpenCL} but not adequately addressed).  The other is that software and hardware programming are in
various aspects still incompatible and therefore not well unified/aligned.
These aspects include time-scale constraints in algorithm development/deployment,
and difficulties/tradeoffs in combining the efficiency offered by hardware with
the flexibility provided by software when implementing algorithmic logic.
These two issues, although orthogonal, are not independent. Realising hardware mechanisms
for network dynamics control, while being able to
customise them at very low latency, requires advances both in expressibility as
well as deployment time-scales of code on hardware. The herein presented work
is a substantial step forward in this respect.

Conventional practice requires the use of
a Hardware Description Language (HDL) such as VHDL~\cite{VHDL} or
Verilog~\cite{verilog}, which in contrast to software programming is a
laborious and time consuming task. While HDL is suitable for
describing sequential and combinational logic, it is very complex and
error-prone when used to implement high-level algorithms (of reasonable
complexity), due to limited expressibility and high-level abstractions at the
level of algorithm behaviour. 
For this reason, complex algorithms are often provided in a
toolbox of manufacturer pre-coded IPCs, which can be used by the hardware programmer
to compose processing pipelines.  For example in the context of traffic
management, Xilinx~Inc.~\cite{ Po06}, Altera~Corp.~\cite{Al14}, and Lattice
Semiconductors~\cite{La14} offer IPCs for multi-level hierarchical queueing,
round robin scheduling, fair queueing, burst equalisation, random early
detection (RED), token/leaky bucket policing, \etc. Yet, this means that, on the customer side,
prototyping, testing, and deploying of new algorithms are still done in software (e.g.~\cite{JeAl14,AlAt08,
DuGi07,MaRa11}), except for a narrow segment of non-novices in hardware
programming (e.g.~\cite{AlKa12,HaDw11}).

To improve the programmer experience by addressing limitations in algorithmic
expressibility, and thereby to bridge the gap between software and hardware
programming (which promotes the widespread adoption of FPGAs), a number of
projects strive to develop language frameworks that raise the level of abstraction from
HDLs (two comprehensive reviews are available in~\cite{Ch12, BaRaSh13}).
Most of these frameworks opt to achieve one or both of the following two
objectives: \lista{}automate code synthesizability (functional verification,
netlist generation, translation and synthesis, mapping to FPGA resource
requirements, place-and-route, timing analysis, bitstream generation) into
something that resembles the compilation process in software languages;
\listb{}formalise ways of mapping a high-level algorithm (behavioural
description) to some low-level description (register-transfer level or digital
circuit). A classification offered by \cite{BaRaSh13} distinguishes five
categories: \lista{}HDL-derived languages enriched with
software engineering features such as object-orientation, type-systems, and
module hierarchies, e.g.,~\cite{Bluespec}; \listb{}C-style language extensions that rely on in-code annotations and confine the programmer
to a small subset of the parent language (e.g. no use of pointers), e.g.~\cite{CaDi09,
CouMo08, Vivado}; \listc{}CUDA/OpenCL-based
frameworks, which
use intermediate data language representations and library IPCs to compile
high-level code into parts that can co-execute on a host CPU and FPGAs (often
supporting dynamic linkage as well), e.g.,~\cite{AlteraOpenCL,JaLa10, PaGu09}; \listd{}modern high-level (often
functional) language-based frameworks, which offer object-orientation, strong
typing, support of polymorphism, and automatic memory management,
e.g.,~\cite{Lime, BaVo12, Kiwi, Esterel}; \liste{}model-based frameworks, which provide graphical representations and rely on
executable specifications to accelerate design and verification, e.g.,~\cite{DaFa06}.

While our work shares similarities with the last category in terms of algorithm
expressibility and representation, there is an important distinction that
differentiates it from all other listed approaches: the time-scale of
program/algorithm deployment and modification.
In the majority of the aforementioned approaches, a high-level expression of an
algorithm is \emph{compiled} offline into HDL code. The netlist (or bitstream)
then needs to be synthesised and ``downloaded'' on the FPGA, requiring an
additional delay. Thereafter, code
modifications require a re-compilation and re-load of the bitstream on the
FPGA (today, a typical delay for downloading a new bitstream into the FPGA is in the
time-scale of \emph{seconds}, thus unacceptable for run-time modifications). These substantial overheads do not exist in our approach, which resembles
more runtime interpretation of programs (rather than off-line compilation).
Program specifications are loaded ``instantaneously'' (in sub-second speeds)
and can be edited while the system is running. This is because CAs have a very simple representation that can be
translated into a set of memory-mapped register values (thus not requiring the
use of slow electronic design automation -- EDA -- tools). Additionally, the inherently
parallel nature of a program in CAs' representation allows different parts of
the algorithm to be modified independently of each other.

Before discussing specifically the chemical middleware abstraction for hardware, we need to 
briefly introduce CAs in general. For the sake of completeness and contextualisation, in the next section, we summarise main principles and concepts (retrievable in \cite{hotnet09}--\nocite{sensorj14, srds12, jsac13,danms13}\cite{ijcnc13,MeTh}).

}


\section{Chemical Algorithms and control of (network) dynamics} 
\label{sec:AC}

CAs (chemical algorithms, or Chemistry-inspired algorithms) refer to a class of stochastic algorithms
whose logic is described and implemented as a chemical reaction network.
Inputs, outputs and internal states are represented by concentrations of
molecular species, and their (mathematical) relationships are represented by
reaction rules. CAs are subject to the kinetics laws of Chemistry (mainly, the
Law of Mass Action and conservation laws), which dominate operations and
influence the behavioural characteristics of the algorithm.  Abiding to
chemical kinetics makes CAs \emph{robust}, \emph{deadlock-free}, and
\emph{analysable}:

\begin{itemize}
\item\emph{Robustness:} CAs are dynamical systems that continuously process
event signals, and are robust to errors or perturbations. Formally, robustness
is the ability of a system, once perturbed from its current trajectory, to find
the attractor (steady-state) that recovers its trajectory.  This happens if,
during the perturbation, the system remains within the basin of attraction.
Steady-state solutions of CAs are attractors with large basins.  In other
words, the system moves in ``small steps'', so that perturbations displace the
system only a small distance from the attractor. By contrast, typical
computational systems that implement discrete time algorithms (network
functions among others) exhibit very small basins of attraction. This means
that the magnitude of perturbations, which the system can absorb without
getting displaced towards a different attractor (error, instability, or
unpredicted state), is rather limited~\cite{St12}.

\item\emph{Deadlock-free operation:} CAs are statistical algorithms with a
deterministic average behaviour. At a ``microscopic'' level, individual
computations (reactions) occur stochastically, independently of each other. 
This means that the algorithm cannot deadlock in some computation or state
(even when the inputs are not synchronised).  At the same time however, the
macroscopic (collective) effects of the entire algorithm have a deterministic
average tendency (as the effects of any single computation are minimal).

\item\emph{Analysability:} CAs are mathematically tractable.  The behaviour of
a CA can be accurately described by a system of equations directly derived from
its (graphical) representation as a reaction network. This is possible because
the internal operation of the CA and the resulting dynamics are governed by the
chemical kinetics laws.  This contrasts the traditional practice of
deriving \emph{a-posteriori} models to approximate the behaviour of already
implemented algorithms.
\end{itemize}

\newtext{
In the last decade, works aimed at establishing and formalising
the chemical metaphor as a computational and programming model in general, e.g.,
\cite{Di05}-\nocite{BaFr05, Pa00, BaDi96,GiMi01}\cite{MaKr11}, and also specifically in networking,
e.g., \cite{sensorj14, danms13,ViCa11, NaGi10}.
}
%


\subsection{Representation of CAs}
\label{sec:CA_representation}

\begin{figure}
 \center
 \begin{minipage}[b]{.52\linewidth}
  \centering
  \centerline{\includegraphics[width =4.5cm]{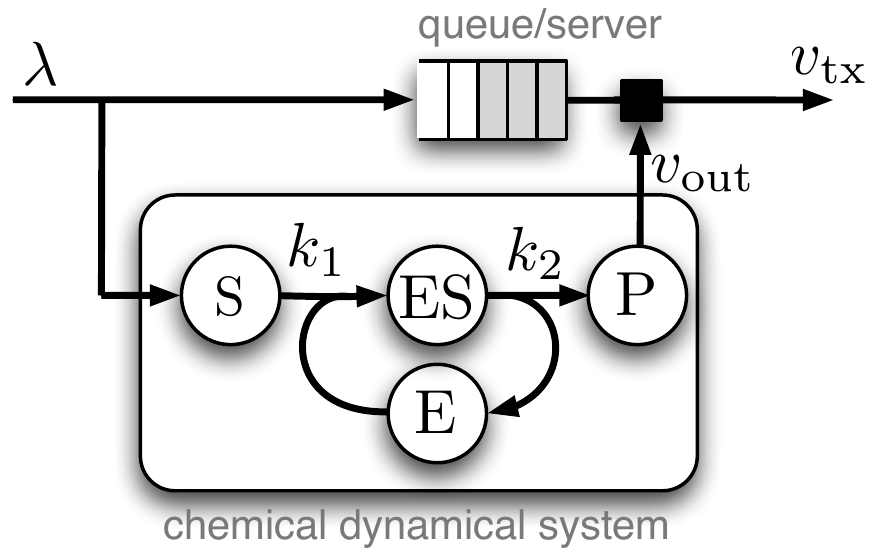}}
  \centerline{(a) System}
 \end{minipage}
 \begin{minipage}[b]{0.45\linewidth}
  \centering
  \begin{subequations}
  \begin{align}
  	 r_{1}:\,\,  & \mol{S}+\mol{E}	 \stackrel{k_1}{\longrightarrow} \mol{ES}	\notag	\\ 
  	 r_{2}:\,\, 	& \mol{ES} 	 \stackrel{k_2}{\longrightarrow}\mol{E} + \mol{P}\notag
  \end{align}
  \vspace{0.3in}
  \end{subequations}
  \centerline{(b) Reactions}
 \end{minipage}
 \caption{\texttt{Rnet1}: The enzymatic reaction network used as a
traffic rate controller (pacing and rate capping). CA's input is
connected to a queue's arrival process and the CA's output controls the queue's service
process.
}
 \label{fig:ChemRateLim}
\end{figure}

Instead of state diagrams or pseudocode that describe a sequential logic, the
logic of CAs is suitably expressed (and visualised) in \emph{drawings} of
chemical reactions among molecular species (e.g., \fig\ref{fig:ChemRateLim}(a),
white rounded-corner square).  The species represent the algorithm's inputs,
outputs, and internal state variables. The reaction network diagram encodes the
parameters that control the behaviour of the system (reaction coefficients and
reactant stoichiometric coefficients). A reaction captures a causal
relationship between the system's state-variables (reactants and products).
Formally, a reaction network (and therefor a CA) is represented by a set
$\mathcal{S}$ of molecular species (variables), and a set $\mathcal{R}$ of
reaction rules of the general form
\begin{equation}\label{eq:reaction} r\in\mathcal R: \quad \sum\limits_{s \in
\mathcal{S}} \alpha_{r,s} s \mathop \to \limits^{k_r} \sum\limits_{s \in
\mathcal{S}} \beta_{r,s} s \;\; , \end{equation}
\noindent which specify how reactant molecules interact to create product
molecules. 
For a reaction $r$, $k_r$ is a constant parameter, known as \emph{reaction
coefficient}, that regulates the relative speed of the reaction ({more}
details later).
Parameter $\alpha_{r,s}$ is the \emph{stoichiometric reactant coefficient},
specifying the number of molecules of a species $s\in\mathcal{S}$
consumed by reaction $r$. Similarly, parameter $\beta_{r,s}$ is the
\emph{stoichiometric product coefficient}, specifying the number of molecules
of a species $s\in\mathcal{S}$ produced by reaction $r$.
In simple words, a reaction rule replaces $\alpha_{r,s}$ amount of molecules
from each species $s\in\mathcal{S}$ with $\beta_{r,s}$ amount of molecules of
each species $s\in\mathcal{S}$ at an average rate controlled by the
$k_r$-coefficient.


A simple example that illustrates a chemical traffic control algorithm is shown
in \fig\ref{fig:ChemRateLim}. As we demonstrate in the following, similar to
the traditional Token Bucket (TB) scheme, this chemical mechanism can be used
to control the service process of a queue and rate cap the outgoing traffic up
to a predefined, adjustable threshold. In contrast with the TB scheme, the
chemical controller allows shaping the outgoing traffic in order to achieve
smooth, burst-free dynamics.\footnote{\newtext{An additional CA that matches exactly the behaviour of the 
TB scheme is discussed in \cite{danms13}.}} The service process, {implemented with a} CA, is
graphically shown in \fig\ref{fig:ChemRateLim}(a) and its logic is formally
described by reactions $r_1$ and $r_2$ in \fig\ref{fig:ChemRateLim}(b). 

For each enqueued packet (or certain amount of bytes), a molecule of species
\mol{S} is created. The dequeueing and transmission of a packet is authorised
by the execution of reaction $r_2$, which implies the production of a \mol{P}
molecule and the consumption of an \mol{ES} molecule.
The production of \mol{ES} molecules in turn is controlled by reaction $r_1$,
and depends on {\mol{S}} molecules (arrivals of packets in the queue) and the
availability of {\mol{E}} molecules, which embody tokens.  Molecules of species
\mol{E} (tokens) are replenished from the separation of \mol{ES} molecules at
the rate at which reaction $r_2$ occurs.
Overall, the {effective} queue service policy is \emph{non work-conserving}:
the queue is not served as fast as possible; its service is instead regulated
by {the relationship between rates of reactions $r_1$ and $r_2$ (as shown in
the next section)}. 

\subsection{Operation and Dynamical aspects}
\label{sec:CA_dynamics}


Dynamics of CAs (when and which reaction is executed) are regulated by the
\emph{Law of Mass Action} (LoMA). The LoMA~\cite{HoJa72} states
that the average rate $v_r(t)$ of occurrence of a chemical reaction $r
\in\mathcal R$ is proportional to its reactant concentrations:\footnote{The
rate value found in \eqref{eq:LoMA} can be regarded as a simplified value
quantifying the propensity $a_r$ of a reaction $r$ to occur
\cite{WoUl04,MeTh}.}
\begin{equation}\label{eq:LoMA}
  v_r(t) = k_r \prod \limits_{s \in \mathcal S} c_s^{\alpha_{r,s}}(t) \;\; ,
\end{equation}
\noindent where $c_s(t)$ denotes the amount of molecules of species $s\in
\mathcal{S}$ at time $t$ ($c_s(t)$ can also be regarded as a {time-continuous},
discrete-valued signal that the system processes), and $k_r$ is the coefficient
that regulates the reaction speed (regulating the relationship between
molecular mass and rate).  
{Reactant concentrations affect the speed of the reaction in a non-linear way,
based on the stoichiometric reactant coefficients -- the sum of reactant
coefficients of a reaction $r$, $\sum_{s \in \mathcal S} \alpha_{r,s}$, is
known as \emph{reaction order}}.

The LoMA couples the state and the dynamics of the system, and plays a key role
in CAs  (as a self-adaptive internal scheduler). For example in the (enzymatic) rate controller in
\fig\ref{fig:ChemRateLim}, the effectiveness of the loop (\mol{E}--\mol{ES}) to
control the transmissions (generation of \mol{P} molecules) stems from the
strict relation that the LoMA imposes between the current state of the system
(how many transmissions have been authorised and how many packets await in the
queue) and the speed along the \mol{E}--\mol{ES} loop. By comparison,
work-conserving scheduling disciplines would cause tokens to loop infinitely
fast, in this way making the mechanism ineffective to shape and limit the
traffic.


The other operational principle behind the automatism of the control loop is
the \emph{mass-conservation} law \cite{SaIn04}, which states that the total sum
of molecule concentrations along a loop remains constant if \lista{}the total
number of molecules consumed by reactions along the loop is equal to the total
number of molecules produced, and \listb{}all concentrations along the loop are
altered only by reactions involved in this or another loop. It follows that, in
the (enzymatic) rate controller in \fig\ref{fig:ChemRateLim}, the number of
tokens $\conc{E}+\conc{ES}=e_0$ is conserved. This limits the maximum number of
\mol{P} molecules that can be generated per second, and thus enforces a rate
cap to the packet transmission.


\subsection{Modeling and Analyzability}
\label{sec:CA_analysis}


In CAs, the dual relationship between system state and dynamics warrants an
exact/accurate mathematical description of the system.  This makes signal- and
control-theory viable tools to analyse the behaviour of the algorithm.


Specifically, the behaviour of each CA is {mathematically expressed as} a fluid
model, i.e., a set of Ordinary Differential Equations (ODEs) of the form 
\begin{equation}\label{genODE} \dot{\vet{c}}(t) = \stoicMat \cdot \vet{v}(\vet
k, \vet{c}(t)).  \end{equation} 
The term on the left-hand side represents the vector of state changes
{(concentration variations)}, whereas the right-hand side {specifies how
reactions effect these changes.} The stoichiometric matrix $\stoicMat$ captures
the {topology of the reaction} network, whereas the reaction rate vector $\vet v$
{encodes the speeds of each reaction}, by {combining} reaction coefficients
$\vet k$ and concentrations~$\vet c$ according to the LoMA in \eqref{eq:LoMA}.
For example, referring back to our rate controller in
\fig\ref{fig:ChemRateLim}, {and given} the reaction set in
\fig\ref{fig:ChemRateLim}(b), the resulting system of ODEs is
\begin{equation} \label{odeChemControl}
	\begin{bmatrix}
 		\dot{\conc{S}}(t) \\ 
		\dot{\conc{E}}(t)\\
		\dot{\conc{ES}}(t) \\ 
		\dot{\conc{P}}(t) \\ 
	\end{bmatrix}=
	\begin{bmatrix}
 		-1 & 0 & 1 \\ 
		-1 & 1  & 0\\
		1 & -1 & 0\\ 
		0 & 1 & 0\\ 
	\end{bmatrix}
	\cdot
	[k_1\conc{S}\conc{E}\;\; \overbrace{k_2\conc{ES}}^{v_\textrm{tx}}\;\; \lambda]^T
\end{equation}
\noindent where the term $k_2\conc{ES}$ reflects the rate of reaction $r_2$ and
thus the dequeueing/transmission rate $v_\textrm{tx}$.

From the developer's perspective, the stoichiometric matrix~$\stoicMat$ provides the means to program any certain CA, and the reaction
coefficient vector $\vet k$ represents the means to calibrate/tune it.  The
concentration vector $\vet{c}(t)$ then represents changes in the CA's state, as
the system evolves over time -- i.e. it is not explicitly controllable.

From \eqref{odeChemControl}, it follows (by solving the homogeneous system for
the steady state) that so long as $\lambda<e_0 k_2$, the concentration \conc{S}
remains stable and the transmission rate $v_ \textrm{tx}^*$ follows the packet
arrival rate $\lambda$ (see \cite{MoSiTR12} for more details).
%
On the other hand, by applying the mass conservation law
($\conc{E}+\conc{ES}=e_0$), one arrives to the Michaelis-Menten (biochemical)
equation: 
\begin{equation*} v_ \textrm{tx}^* = k_2 \conc{ES} = \overbrace{ e_0
k_2}^{v_\textrm{max}}\frac{ \conc{S}}{(k_2/k_1)+\conc{S}}\;\; , \end{equation*} 
from which it draws that when $\lambda>e_0 k_2$, and thus when \conc{S} grows
without bounds, the transmission rate $v_\textrm{tx}^*$ grows asymptotically
towards the rate cap of $v_\textrm{max}$, prescribed in the product of the
terms $e_0$ and $k_2$.  The ratio $k_2/k_1$ controls how fast the rate limit is
enforced.

From the transient/sensitivity analysis in \cite{MoSiTR12}, it stems that the
control algorithm has a low-pass filtering behaviour. The cut-off frequency is
directly controllable through $k_2$-coefficient (i.e., higher $k_2$ values lead
to higher cut-off frequencies -- the outgoing traffic from the system is more
bursty).

Apart from a fluid-model analysis, one may also study a CA at the
microscopic level with queueing theory (and thus complement
control/signal theory applied at the macroscopic level) \cite{MeTh}. A
molecular species represents a (virtual) queue and thus, a chemical reaction
diagram depicts a network of interacting queues. The queue service process obeys chemical kinetics and has a specific
mathematical form, which manifests in the system model description and
analysis. As a consequence,
the relationship between arrival and departure process is not only
predictable but also exploitable as a design feature in the engineering of the
algorithm. Based on this, we are able to design a system ``by queue
interactions'' (by applying reaction rules), and prescribe how departure
processes of queues modulate each other.


\section{Deploying CAs on FPGA technology}\label{sec:FPGAimplHW}
\newtext{
After having discussed CAs in general, we now exploit the introduced concepts to describe and implement 
a generic programmable hardware platform, particularly suitable for Field Programmable Gate Arrays (FPGAs) technology.
}

FPGAs are the most preferred platform for
introducing new network functions close to the hardware. The reasons are the
fast time-to-market, the low-cost realisation, the extended re-programmability
(compared to Application Specific Integrated Circuits, ASICs), and the high amount of
available logic resources (compared to Complex Programmable Logic Devices, CPLDs).

To enable a generic programmable deployment of CAs on FPGAs, we sought to provide a
``chemical engine" abstraction. This middleware abstraction serves the
following two key purposes. On one hand, it hides low-level hardware
description in ``chemical" primitives, which leverage a high-level description of
CAs through the reaction network representation. On the other hand, it
considerably reduces the programming time of CAs based on a two-level
configuration process.

\newtext{
At low level (level-1), the construction of a chemical engine on the FPGA
creates chemical resources and an execution environment. This requires a
``traditional'' slow field-programming process involving synthesis of HDL code
and bitstream generation, which is acceptable as a system initialisation (e.g.
boot-time) task. The generated execution environment provides all the
background functionality for setting up a CA and embodies the chemical kinetics
for running it. At high level (level-2), the actual programming of CAs is
effected as a configuration task that allocates part of these resources and
connects them in the corresponding reaction network. These resources can be
re-allocated or modified at any time (through a new level-2 configuration) to
implement another CA. That is, the level-2 configuration is the essence of the
\emph{fast runtime} programmability of CAs.
}

Formally, the instantiation of a CA (level-2) within the chemical engine
(level-1) completes the implementation of a so-called \emph{Artificial
Chemistry} \cite{DiZi01}
$\AC{}=\{ \{ \mathcal{S}\},\{ \mathcal{R}\}, \mathcal{A}\}$.
The level-1 generated, chemical engine provides the LoMA reaction scheduling
logic $\mathcal{A}$ in the execution environment. At level-2, configuration
provides the structural information (species set $\{ \mathcal{S}\}$ and the
reaction set $\{\mathcal{R}\}$) for any CA.

In the following, we present in more detail the key components of this chemical
middleware, and discuss its implementation in an FPGA device by Xilinx.


\subsection{Chemical Engine Middleware -- Platform Overview} 

\begin{figure}[!t]
\centering
\includegraphics[scale=0.22]{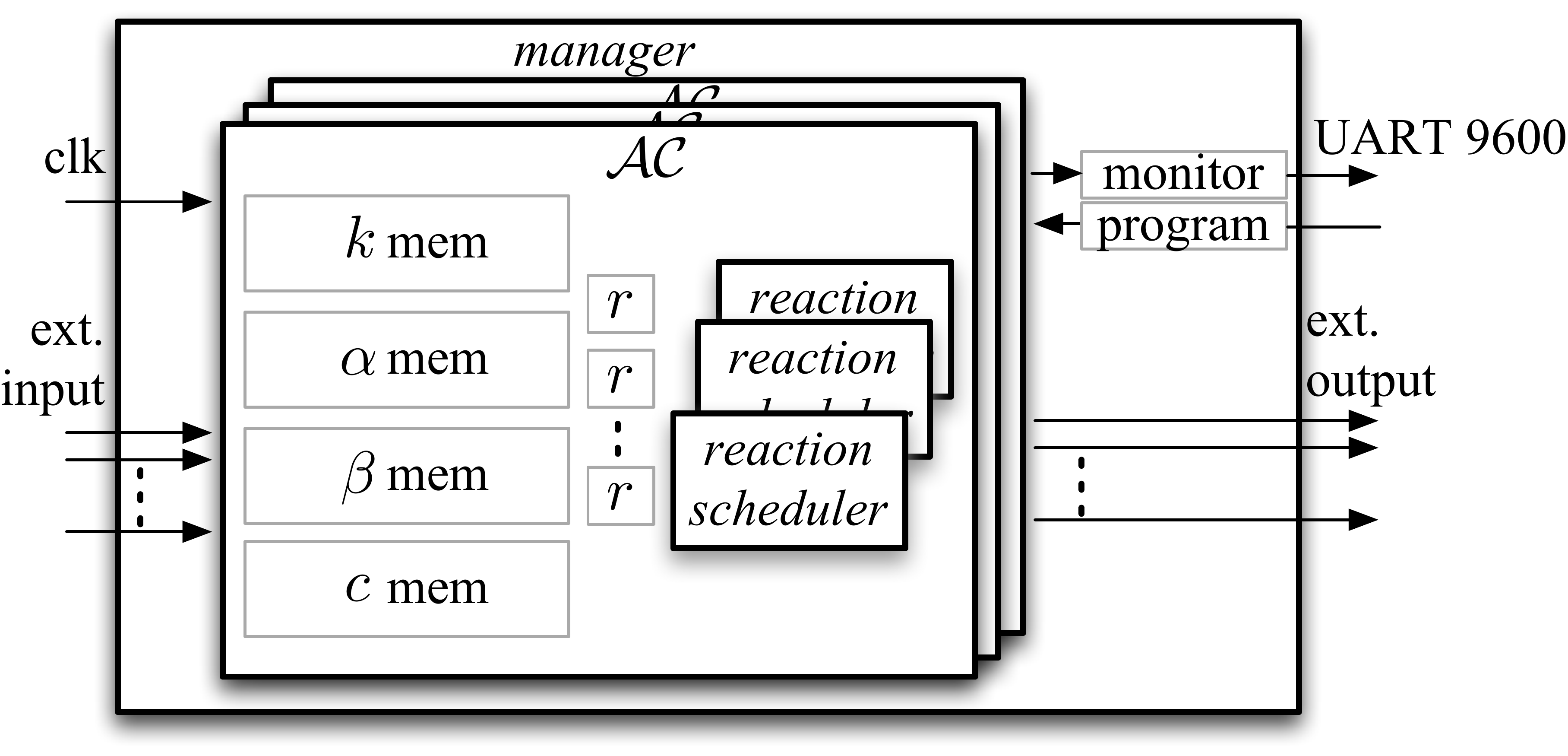}
\caption{
{Block-diagram illustrating the main components of the chemical middleware 
platform for programming CAs on FPGA hardware.}
}
\label{fig:HW_atGlance}
\end{figure}

The key building blocks (operational modules and functional structures) of the
engineered chemical middleware platform are shown in the block diagram of
\fig\ref{fig:HW_atGlance}. The runtime operation is divided across three main
nested modules: 
\lista{}the \texttt{manager} module, 
\listb{}the \AC{} module that implements one or more
\emph{chemical engines} as part of the CA execution environment, and \listc{} the
\texttt{reaction-scheduler} module (LoMA core) that implements the reaction algorithm
\algAC{} and schedules reactions for execution.

Specifically, the \texttt{manager} module may serve simultaneously (taking
advantage of the hardware parallelisation) more than one \AC{} modules, each
hosting a separate CA. It handles the I/O for each \AC{} module by mapping
input and output signals (events such as packet arrivals) to specific species
of a CA. It also facilitates programming of CA and monitoring of its state by logging
periodically the concentration values of selected species. 

An \AC{} module represents the principal component of the implementation of CAs
in hardware. It hosts in memory the functional data structures (tables) for the
{structural} representation of a CA -- i.e., species concentrations,
stoichiometric reactant and product coefficients, and reaction coefficients.
Values in these structures, which are runtime accessible, {provide} the
inputs to the hardware logic circuitry embedded in the \AC{} module, which
implements the addressing mechanism to inter-wire the CA at runtime.  For
example, the values of the stoichiometric memories $\alpha$ and $\beta$ decode
the addresses of reactant and product concentrations of each reaction.
Similarly, the reaction coefficients stored in the $k$ memory affect the
computation of \emph{next reaction time}.

The \texttt{reaction-scheduler} module (LoMA-core) computes the propensity of a reaction
from its reactant concentrations and from the reaction coefficient, and
produces as output the time at which a reaction should be executed.


\subsection{Reaction Network -- CA topology}

The approach we have adopted to enable runtime programmability of CAs consists
of two phases: First, during the hardware programming of the FPGA, a ``large
enough'' grid of chemical resources (in hardware logic circuitry) are reserved.
Second, at runtime, these resources are allocated under user-defined
configurations to instantiate different CAs.  \newtext{This involves merely the 
setting of values in memory-mapped registers on the FPGA.}
To do this, the user defines a number of species, the initial values for their
concentrations, and a set of reactions with their coefficients.  This
information ``topologically interconnects'' a
CA and configures the dynamics of its execution (when reactions occur). In this
section, we describe how the reaction network topology is fleshed out on
hardware following a CA configuration. In the next section, we explain how the
dynamics of reactions {are} orchestrated.

%
Each species is implemented as a register made up of a chain of flip-flops,
whose number determines the maximum value (as a power of 2) that a
concentration can assume.  
%
%
Reaction rule definitions, on the other hand, provide information about
which species engage as reactants, which as products, and in what
quantities (respective stoichiometric coefficients).

\begin{figure}[t]
\centering
\includegraphics[scale=0.126]{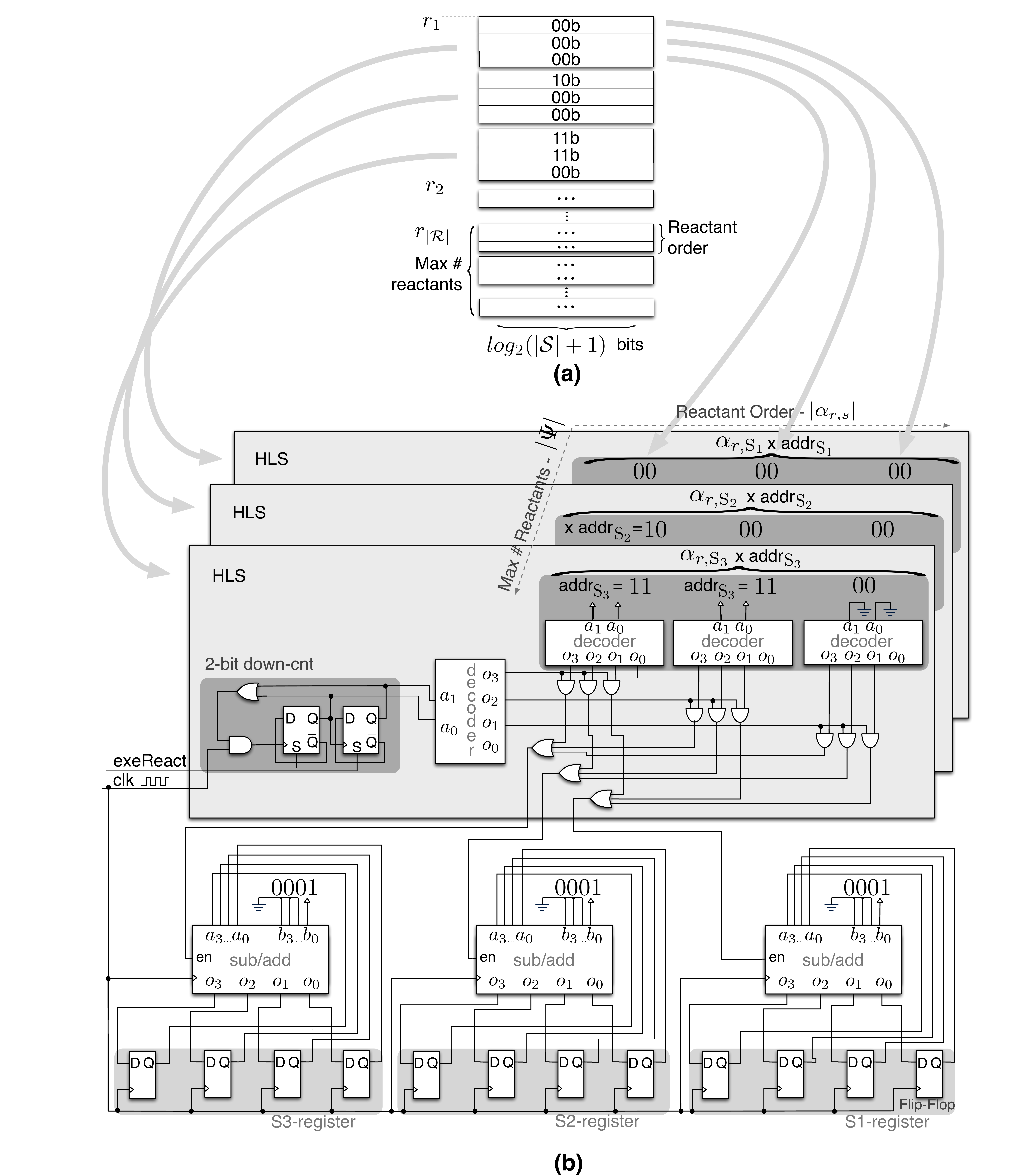}
\caption{
{Addressing logic for updating reactant concentrations (analogous for product
concentrations). (a) 3D organisation of stoichiometric memory of reactants:
reaction $\rightarrow$ reactant number $\rightarrow$ reactant order.  (b)
Circuitry schematic related to the stoichiometric memory (example for maximum 3
species, 3 reactants per reaction, up to 3rd order reactants).}
}
\label{fig:stoichROM}
\end{figure}

The stoichiometric information of reactants and products is divided in two
respective 3D structures, whose top-level index corresponds to each
individual reaction (\fig\ref{fig:stoichROM}(a)). The size of these tables
(programmed on the FPGA) defines the maximum resource allocation available to
the user for configuring CAs at runtime. The information stored therein is used
to actuate the \emph{addressing} and \emph{computing} logic components of the
chemical engine, in order to update the species concentrations whenever a
reaction takes place.  In the following, we will confine our discussion to the
operations involving the reactants only; analogous description holds for
product species with the sole difference that logic elements for addition
replace those for subtraction (\fig\ref{fig:stoichROM}(b)).

The stoichiometric table $\alpha$-\texttt{mem} of reactants (\fig\ref{fig:stoichROM}(a)) is
dimensioned by reaction (1D), by reactant (2D), and by reactant
coefficient/order counter (3D).  That is, within each indexed reaction record
at the 1st-level, there is a sub-indexing of a maximum number of independent
reactants. In turn, within each indexed reactant at the 2nd-level there is
another sub-indexing of records that contains either a reactant species'
address (\emph{active}) or null (\emph{inactive}). If a reactant
species' address is duplicate in several 3rd-level records, these records
enumerate the respective reactant coefficient (reactant order). At least one
active record implies a 1st-order reactant, which activates the nested
(2nd-level) indexed reactant position and in turn the outermost (1st-level)
indexed reaction record.


As seen in \fig\ref{fig:stoichROM}, the structure of the stoichiometric tables
reflects the fact that the processing for each indexed reactant (2nd-level)
takes place in a separate \emph{Hardware Logic Slice} (HLS) -- vertical
arrangement.  HLSs can be engaged in parallel in computations of the CA, such
that reactions that involve 1st-order reactants (e.g.,
$\mol{S}_1+\mol{S}_2+\mol{S}_3 \rightarrow \ldots$) can be processed in
parallel in a single step. 

The number of \emph{active} address-records at the 3rd-level (encoding the
reactant order) enumerates how many processing steps are required to complete
the update of reactant state, during the execution of the reaction. 3rd-level
address-records directly index a respective number of decoder elements within
each reactant's HLS. Each decoder is activated in sequence through a step-down
counter.  The address stored in each address record of the stoichiometric table
is input to the decoder so to actuate a subtraction operation on the respective
species concentration.  As a result of this process, a reaction of the sort
$3\mol{S}_1 \rightarrow \ldots$ is computed in a number of steps that reflects
the reactant order $(\mol{S}_1) + (\mol{S}_1) + (\mol{S}_1) \rightarrow \ldots$
({where} each parenthesis {pair denotes} a single processing step). 


The \emph{maximum} number of indexable reactions (1D), reactants-per-reaction (2D) and
reactant order (3D) records needs to be fixed at the time of programming the
FPGA. For example, the chemical engine encoding the addressing logic of
\fig\ref{fig:stoichROM}(b) refers to a resource
reservation (maximum allocations) for 3 species with concentration size up to
15 molecules, and one indexable reaction with at most 3 reactants/products per
reaction, and of up to 3rd order each.  
For 3 species, 2-bit addresses are needed to resolve access to their registers
(S3, S2, S1), each of which is 4-bit wide (number of flip-flops in each
register){, and thus holding} concentration size values $\leq$ 15. 
The corresponding reactant stoichiometric table (see \fig\ref{fig:stoichROM}(a)
for reaction $r_1$) indexes reactions (1st-level), each of which {sub-}indexes
maximum 3 reactants (2nd-level), each in turn {sub-indexes} maximum 3 address
records (3rd-level) for enumerating the order (maximum of the 3rd order) of a
reactant.

For a configured CA that involves a reaction of the form $2\mol{S}_3 +
\mol{S}_2 \rightarrow \ldots$ (\fig\ref{fig:stoichROM}), the two reactants
$\mol{S}_2$ and $\mol{S}_3$ occupy two 2nd-level records (out of the three
available). The one corresponding to $\mol{S}_3$, which is a 2nd-order
reactant, has two 3rd-level records (out of three available) filled with the
species address \texttt{11b} of the $\mol{S}_3$ register. By analogy, for the
1st-order reactant $\mol{S}_2$, only one 3rd-level record (out of three
available) is filled with the species address \texttt{10b} (refer to
\fig\ref{fig:stoichROM}(a)).

When the reaction executes, through the \texttt{exeReact}-signal, each of
its reactant species is processed at a different HLS, allowing for their
parallel computations. I.e., $\mol{S}_3$ will be processed at the frontmost
HLS, $\mol{S}_2$ at the next, while the last HLS will remain unused since there
are only two reactants.
Within each HLS, i.e. for each reactant, the 2-bit species address stored in
each 3rd-level record of the stoichiometric table is input to one correspondent
decoder. For reactant $\mol{S}_3$, its address \texttt{11b} appears in the
inputs of two of the three decoders.
The output of each decoder is read in subsequent steps of the step-down counter
and activates (\texttt{EN}-input) a subtracter that decrements by 1
molecule (in every step) the contents of the respective species register.  In
effect, this reduces the concentration of $\mol{S}_3$ by 2 in two steps, and
respectively the concentration of $\mol{S}_2$ by 1 in one step.  Overall, the
discussed hardware logic computes
$2\mol{S}_3 + \mol{S}_2 \rightarrow \ldots$~, as $(\mol{S}_3 + \mol{S}_2) +
(\mol{S}_3) \rightarrow \ldots$~.


\subsection{Reaction Scheduling}

Reactions are executed in real-time according to a time-schedule that {abides
to the LoMA} (see \sect\ref{sec:AC}). Computing the reaction-times schedule is
the most costly operation, in terms of hardware logic.
%
After a reaction has fired, and the update of species concentrations for
reactants and products has been performed, a \emph{{next} reaction-time}
computation is triggered for each dependent reaction (i.e., all reactions whose
reactant concentrations have been modified).
For a reaction $r$, this requires to compute the propensity, i.e., the product
of reactants' concentrations $c_s^{\alpha_{r,s}}$ and the reaction
coefficient~$k_r$, see \eqref{eq:LoMA}.  The reaction coefficients $\vet k$
are stored in a separate bank of registers.

\begin{figure}[!t] \centering
\includegraphics[scale=0.135]{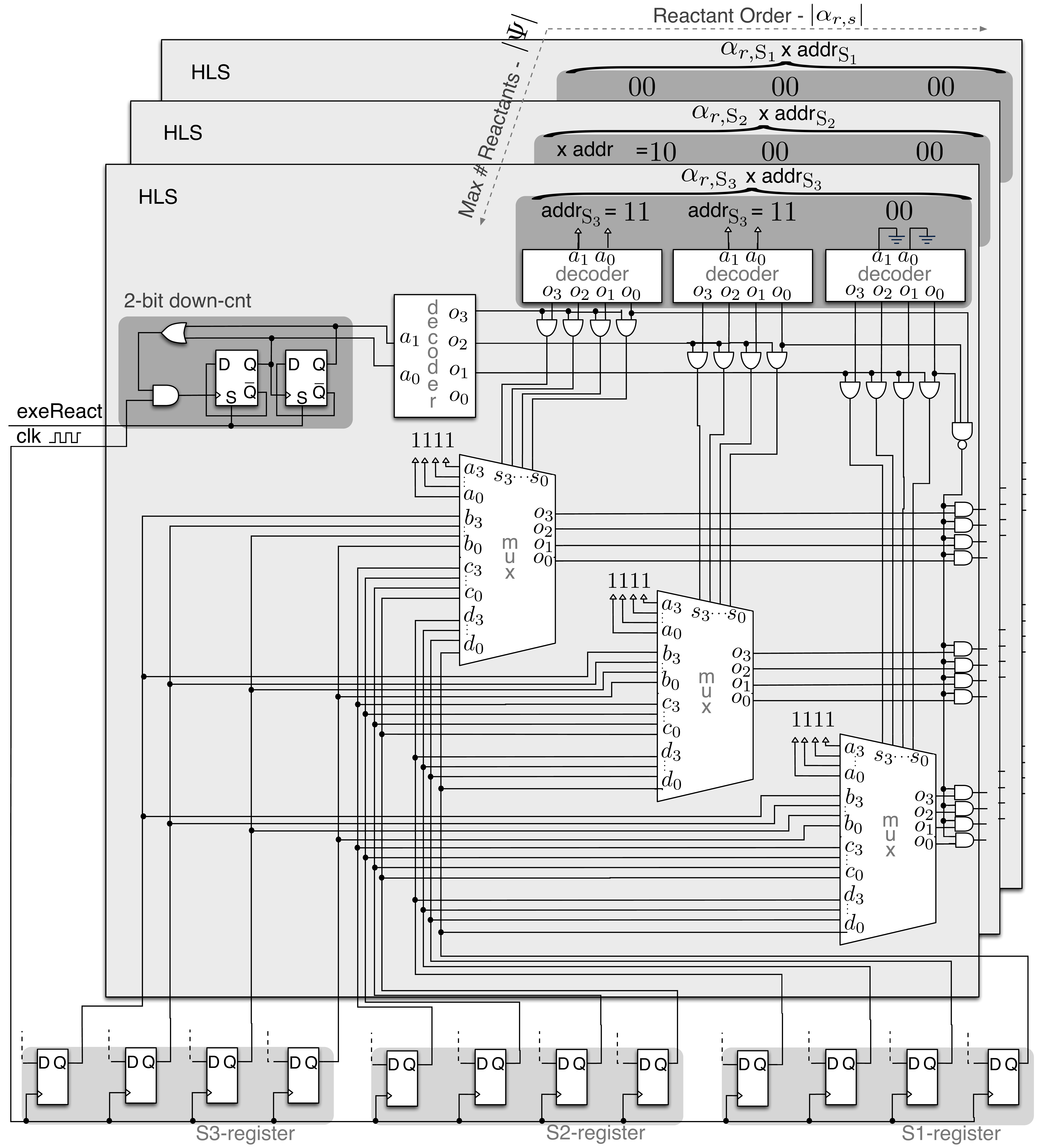} 
\caption{
{Addressing logic for selecting concentrations to compute reaction
propensities (example for maximum
3 species, 3 reactants per reaction, up to 3rd order reactants).}
}
\label{fig:stoichForTime} 
\end{figure}

To select the (reactant) species needed for {computing the propensity of each
dependent equation,} we use the hardware logic circuit shown in
\fig\ref{fig:stoichForTime}.
{Just like with the addressing logic for updating the concentrations in the
previous section,} we rely on the information from the reactant stoichiometric
table to index across HLSs and decoders. {However,} in this case, the output of
each decoder selects inputs of a chained-up multiplexer. At every
step of the counter, one multiplexer outputs the value of the decoded species
register (for $s_3\dots
s_0=\texttt{1000}$ it forwards the value of the $S3$-register, for $s_3\dots
s_0=\texttt{0100}$ the value of the $S2$-register, and for $s_3\dots
s_0=\texttt{0010}$ the value of the $S1$-register), or the fixed value
$\texttt{1111}$, for the identity element of the multiplication.

Outputs from each HLS {will contribute to the computation of} the power of each
reactant concentration $c_s^{\alpha_{r,s}}$ (e.g. $c_{\textrm S_3}^2$), while
the combination of the outputs across HLSs {will} contribute to {the
computation of} the product of reactants' terms $\prod_{s \in \mathcal S}
c_s^{\alpha_{r,s}}$ (e.g.  $c_{\textrm S_3}^2  c_{\textrm S_2}$).
To {complete the computation of} the propensity, these values alongside the
reaction coefficient $k_r$, are input to a logic module for multiplication.
Depending on the required trade-off {between} logic density and computation
speed, this operation can be performed by a single multiplier in as many as $|\Psi| \times
|\alpha|$ time steps ($|\Psi|$ being the maximum possible number of reactants,
and counting in the additional multiplication by $k_r$), or by up to $|\Psi|$
parallel scaled-multipliers in as {little} as $|\alpha|$ time steps.

The computation of the new time schedule thereafter requires \lista{}to compute
the reciprocal of the propensity value in order to calculate the next reaction
time for the reaction that was just executed, and \listb{}{possibly to rescale
the old propensity value for all dependent reactions, so as to update their
time schedules according to new reactant concentrations.}

This process can be speeded up by parallelising propensity and reaction-time
computations by means of separate \texttt{reaction-scheduler} modules. The
number of \texttt{reaction-scheduler} modules (ranging from $1$ to $|\mathcal
R|$; from one per \AC{} up to one per reaction) represents the tradeoff between
speed and logic utilisation.


\subsection{Realisation on Xilinx Spartan-6 FPGA Family}\label{sec:FPGAimplBoard}

We have realised the middleware framework for CAs and the chemical engine
abstraction discussed so far, on a relatively small, low-cost FPGA device: the
Xilinx Spartan-6 XC6SLX9 (see~\cite{Xi11FPGA} for a general overview on its
features) mounted on the Avnet Spartan-6 LX9 MicroBoard~\cite{Av11}.

To perform computations required in the \texttt{reaction-scheduler} module, we
have used the Xilinx single-precision floating-point IPC~\cite{Xi12} (compliant
with IEEE-754 Standard~\cite{IE08}), which gives us a wide dynamic range
($\sim{\pm}2^{127}$) and a good resolution (${\sim}2^{-23}$) for representing
{floating point} variables during the reaction-time schedule calculations.


For the experiments described in the following section, we have programmed the
chemical middleware platform on the XC6SLX9 FPGA, with resource specifications as shown in
Table~\ref{tab1} and Table~\ref{tab2}.\footnote{\newtext{Source
VHDL codes can be retrieved from the URL http://cn.cs.unibas.ch/projects/HWAC/.}}  A single
chemical engine hosts up to 255 species {and} 8 reactions of the {8th} order,
with up to 8 reactants {and} products.  For most of the practical applications
we have dealt with, reactants/products are of 1st or {2nd} order, and reactions
rarely involve more than 3-4 reactants and 1-2 products each.
The $c$-\texttt{mem}, storing {species} {concentrations}, is 16-bit wide
allowing concentrations to grow up to $2^{16} - 1$.  Its locations are
initialised to $00...0b$, except for the first (reserved) position set to
$00...1b$.  {Concentrations that are connected to input/output events are
updated in batch quantities according to a molecules-per-event ratio.} {The
$\alpha$-\texttt{mem} and $\beta$-\texttt{mem} store stoichiometric
information.} The $k$-\texttt{mem} stores single-precision floating-point
values (32 bits) of reaction coefficients.
\begin{table}[t]
\centering
{\footnotesize
\begin{tabular}{c l l}
\hline
parameter           & value  & description \\
\hline

$|\mathcal{R}|$  & 8      & max number of reactions \\

$|\Psi|$         &{8}      & max number of reactants/products \\

$|\mathcal{S}|$  & 255    & max number of species \\

$|C|$            & 16bit  & max concentration value/size ($2^{|C|} - 1$) \\

$|\alpha|$       &{8}      & max reactant stoichiometric coefficient value\\

$|\beta|$        &{8}      & max product stoichiometric coefficient value\\

$|k|$            &{32}bit  & reaction coeff. size (single-precision floating point)\\
\hline
\end{tabular}
}
\caption{
{Chemical middleware platform resource reservation programmed on the
XC6SLX9 FPGA for our experiments.}
}
\label{tab1}
\end{table} 
\begin{table}[t]
\centering
{\footnotesize
\begin{tabular}{c r l}
\hline
table & &size \\
\hline

\cmem  & $|\mathcal{S}|$ pos &\texttt{x}\;\; $|C|$ bit \\

\amem  & ($|\mathcal{R}|$ \texttt{x} $|\Psi|$ \texttt{x} $|\alpha|$) pos &\texttt{x}\;\; $log_2(|\mathcal{S}|$) bit \\

\bmem  & ($|\mathcal{R}|$ \texttt{x} $|\Psi|$ \texttt{x} $|\beta|$) pos &\texttt{x}\;\; $log_2(|\mathcal{S}|$) bit \\

\kmem  & $|\mathcal{R}|$ pos &\texttt{x}\;\; $|k|$ bit \\
\hline
\end{tabular}
}
\caption{
{Capacities of CA memories}
}
\label{tab2}
 \end{table} 
\begin{figure}[t]
\centering
\includegraphics[scale=0.8]{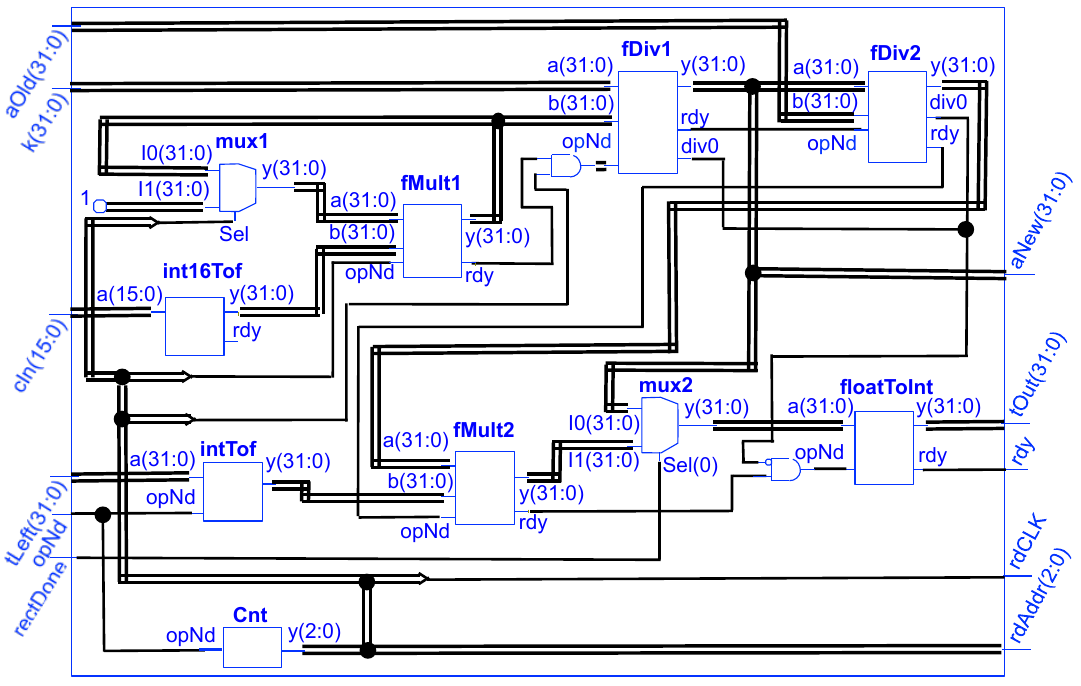}
\caption{
{Schematic of the \texttt{reaction-scheduler} module. It integrates 2 int-to-float and 1 float-to-int converters, 2 multipliers, 2
divisors, 2 multiplexers, and 1 counter.}
}
\label{fig:scheme}
\end{figure}

\fig\ref{fig:scheme} shows the hardware logic layout for computing the
reaction-time schedules. The multiplication of the reactant concentrations is
performed iteratively by a single floating-point multiplier only, because of
restrictions in the amount of available logic on the XC6SLX9 chip. This
means that we loose in parallelisation because we have limited the number of
floating-point operations to two multiplications and two divisions for each
time schedule computation. 



\section{Evaluation}\label{sec:expHW}

In this section, we provide an evaluation of the chemical engine middleware
platform based on our prototype implementation in the XILINX Spartan6 FPGA
family. The objectives of this evaluation are 

\begin{itemize}
\item to demonstrate the runtime programability on hardware of (chemical)
algorithms to control network traffic dynamics 
\item to quantify the performance gains enabled by running CAs on hardware.
\end{itemize} 

\noindent To keep the discussion focused on these evaluation objectives,
{and avoid introducing new algorithms,} we present the experiments with
the exemplary CA that has been used in our discussions until now.
In \sect\ref{sec:discussion}, we briefly report on experiments with
other CAs and their applications.


\subsection{Experiment setup}\label{sec:expSetup}

We used CAs to control {the service process of
the egress queue of a standard Linux host (Linux, Kernel 3.8.6), and thereby
shape its outgoing traffic.} 

We employ the \texttt{tc} tool to isolate a class of traffic in a separate FIFO
queue. The arrival process of that queue provided the input for the CA: for
each enqueued packet, an amount of molecules corresponding to the number of
bytes in the packet was added to an input species \mol{S} in the chemical
engine. On the other end, an output species \mol{P} was ``connected" to the
service process of the queue: for each \mol{P}-molecule produced, a fixed
number of bytes were allowed to leave the queue; when there were enough
molecules to match the byte-size of the packet at the front of the queue, the
packet was dequeued and transmitted. In both cases, the molecules-to-bytes
ratio was kept fixed at 1 mol/KB. 

To interface the FPGA (LX9) board, where the chemical engine lies, with the
queue management subsystem of the linux kernel, we used the \texttt{Parapin}
kernel module. \texttt{Parapin} module allows the use of the PC's parallel port
as a custom I/O interface (i.e., allows handling interrupts at the port pins,
and accessing directly the parallel port registers). We then wired one of LX9's
I/O connectors to the parallel port of the PC. {With such an interfacing, it
was possible to produce/process} interrupts every 100~ns.

The results shown in the graphs that follow concern UDP traffic produced with
the \texttt{iperf} tool (client side running on the controlled node).  {We have
not included measurements with TCP traffic because there, effects of the CA
controller are coupled with TCP's control-loop behaviour, and thus are not easy
to evaluate.}

\begin{figure}[t]
\centering
\includegraphics[scale=0.25]{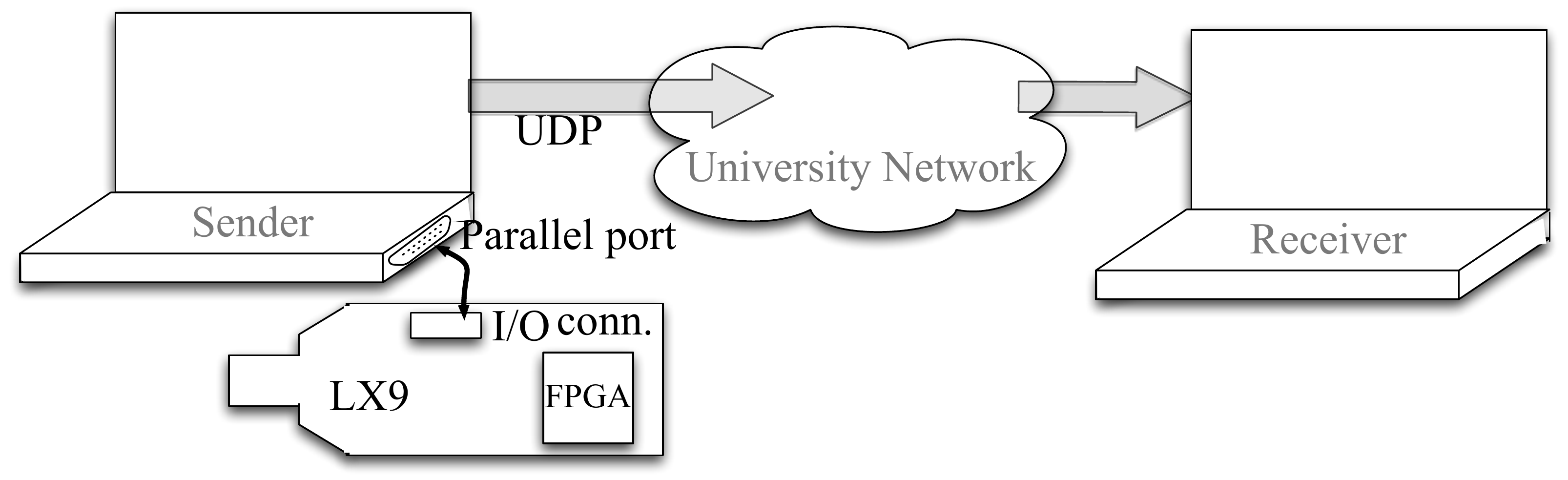}
\caption[Traffic-rate-controlling experiment -- setup]{
{Experiment setup to rate control  PC's egress traffic by means of CAs. The
FPGA hosting the chemical engine was connected to the parallel interface of the
sender host for facilitating the signalling between the CA and the
queue-management subsystem of the linux kernel.}
}
\label{fig:HW_enzy_setup}
\end{figure}

{\fig\ref{fig:HW_enzy_setup} shows the host-to-host topology of the experiment, 
over the high-speed switched network of the university.}

\subsection{Runtime {programmability on} hardware}\label{sec:expProgrammability}

\begin{figure}[t]
 \center
 \begin{minipage}[b]{.52\linewidth}
  \centering
  \centerline{\includegraphics[width =4.5cm]{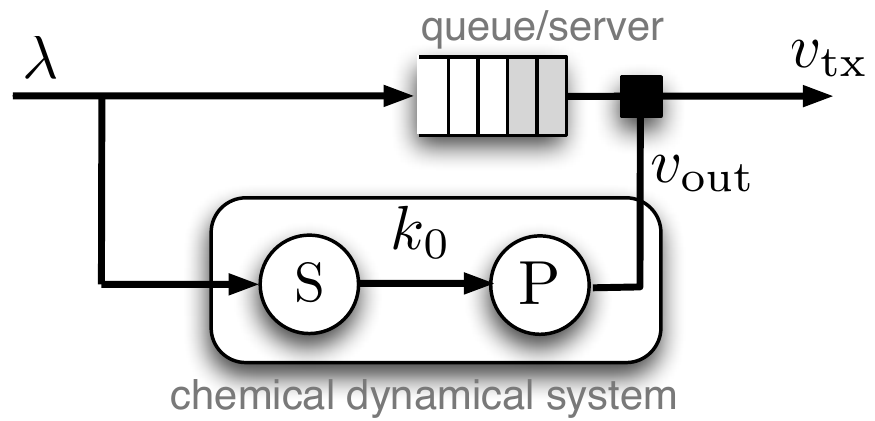}}
  \centerline{(a) System}
 \end{minipage}
 \begin{minipage}[b]{0.45\linewidth}
  \centering
  \begin{equation}
  	 r_\textrm{0}:\,\,   \mol{S}	 \stackrel{k_0}{\longrightarrow} \mol{P}	\notag
  \vspace{0.3in}
  \end{equation}
  \centerline{(b) Reaction}
 \end{minipage}
 \caption{
 {\texttt{Rnet2}: Simple reaction network enforcing the LoMA (eq.\eqref{eq:LoMA}) as a queue service process, so as to implement a traffic pacer.}
}
 \label{fig:LoMARateLim}
\end{figure}

To test and demonstrate the runtime programmability of CAs on {the
FPGA-embedded} chemical engine, we first instantiated in the system a simple
CA (\fig\ref{fig:LoMARateLim}) that paces packet transmissions by a variable
time delay. 
{The simple reaction network essentially imposes the LoMA \eqref{eq:LoMA} as a queue
service policy.}
The CA ``program'' is essentially the following
{reaction network} specification:

\begin{figure}[H]
 \vspace{-0.15in}
\begin{minipage}[b]{0.99\linewidth}
\footnotesize
 \centering
\begin{equation*}\label{eq:HWlomaAC}
\mathcal{S}=\{\mol{S}, \mol{P}\}, 
\,\,\,\,\mathcal{R}=
\{r_0\}
\end{equation*} 
 \begin{equation*}
 \begin{tabular}{ c}
 $c_\mol{S}^0=c_\mol{P}^0=0,\,\,\,\,\,k_0=20$~s$^{-1}$\\\\
 \end{tabular}
\end{equation*}
\end{minipage}
\label{fig:LoMAHWConfig}
 \vspace{-0.15in}
\end{figure}

\begin{figure}[t]
\centering
\includegraphics[scale= 0.5]{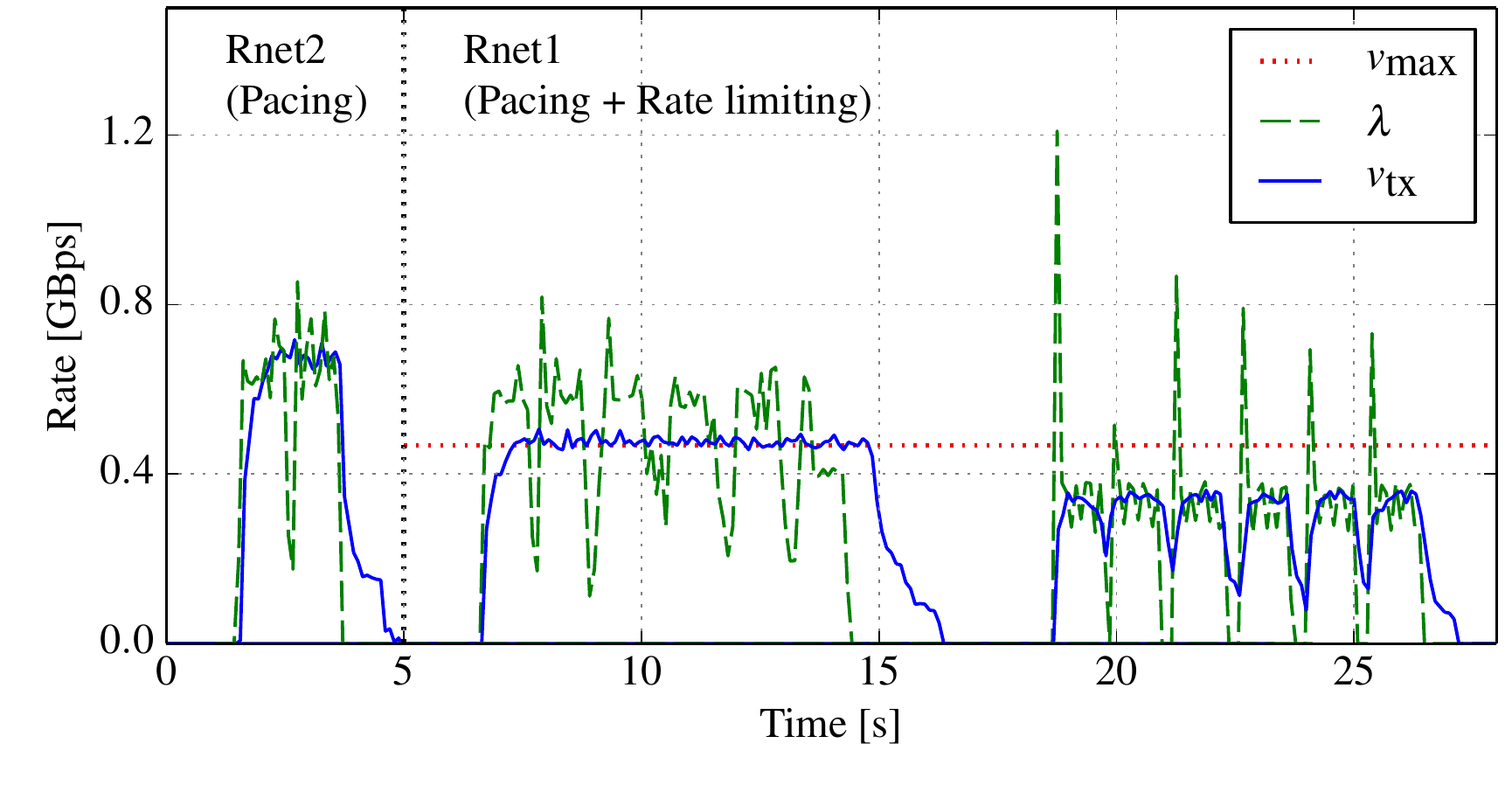}
\caption{
{Traffic shaping effects of 2 programmed CAs: Between
t=[0-5s] the chemical engine was programmed with \texttt{Rnet2}, then between
t=[5-27s] the chemical engine was re-programmed with \texttt{Rnet1}. $\lambda$
is the input rate (load presented by the network layer), $v_\textrm{max}$ is
the rate limit set by \texttt{Rnet1}, $v_\textrm{tx}$ is the output rate
(actual transmissions authorised by the CAs).}
}
\label{fig:expHWEnzyControl}
\end{figure}

In the first $5~s$ of the experiment (see \fig\ref{fig:expHWEnzyControl}), the
output rate followed the average load of the queue. At the same time the
burstiness of the arrival process was smoothed out (filtering of high frequency
components). The cut-off frequency for such a filtering was set via
$k_0$-coefficient (=$20$~s$^{-1}$).

After $t=5~s$, we re-programmed the \AC{} with \texttt{Rnet1}
(enzymatic rate controller -- see {\fig\ref{fig:ChemRateLim}}), by loading the
following CA specification:

\begin{figure}[H]
 \vspace{-0.15in}
\begin{minipage}[b]{0.99\linewidth}
\footnotesize
 \centering
\begin{equation*}
\mathcal{S}=
\{\mol{S},\mol{E},\mol{ES},\mol{P}\}, \,\,\,\,
\mathcal{R}=\{r_1,r_2\}
\end{equation*} 
 \begin{equation*}
 \begin{tabular}{ c}
 $c_\mol{E}^0=25Kmol,\,\,\,c_\mol{S}^0=c_\mol{ES}^0=c_\mol{P}^0=0,\,\,\,\,\,k_1=1$~(mol$\cdot$s)$^{-1}$,$\,\,\,\,\,k_2=20$~s$^{-1}$\\\\
 \end{tabular}
\end{equation*}
\end{minipage}
\label{fig:enzyHWConfig}
 \vspace{-0.15in}
\end{figure}

\noindent{Setting $e_0=c_\mol{E}^0=25Kmol$ and $k_2=20$~s$^{-1}$ fixed}
the rate cap at 0.5~Gbps.

In $t=[6.5,14~s]$ (see \fig\ref{fig:expHWEnzyControl}), a new
round of UDP transmissions increased the load above the predefined rate cap.
The output rate ramped-up to the cap rate, and remained at that limit
until the transmission ended. A third round of UDP transmission started at
time $t=19$~s. The load still had mixed high-frequency and low-frequency bursts,
but this time did not exceed the rate cap. The CA worked as a pacer: the
transmission rate followed closely the slow fluctuations of the arrival rate,
but very high-frequencies were filtered out.

\begin{figure}[t]
\centering
\includegraphics[scale= 0.5]{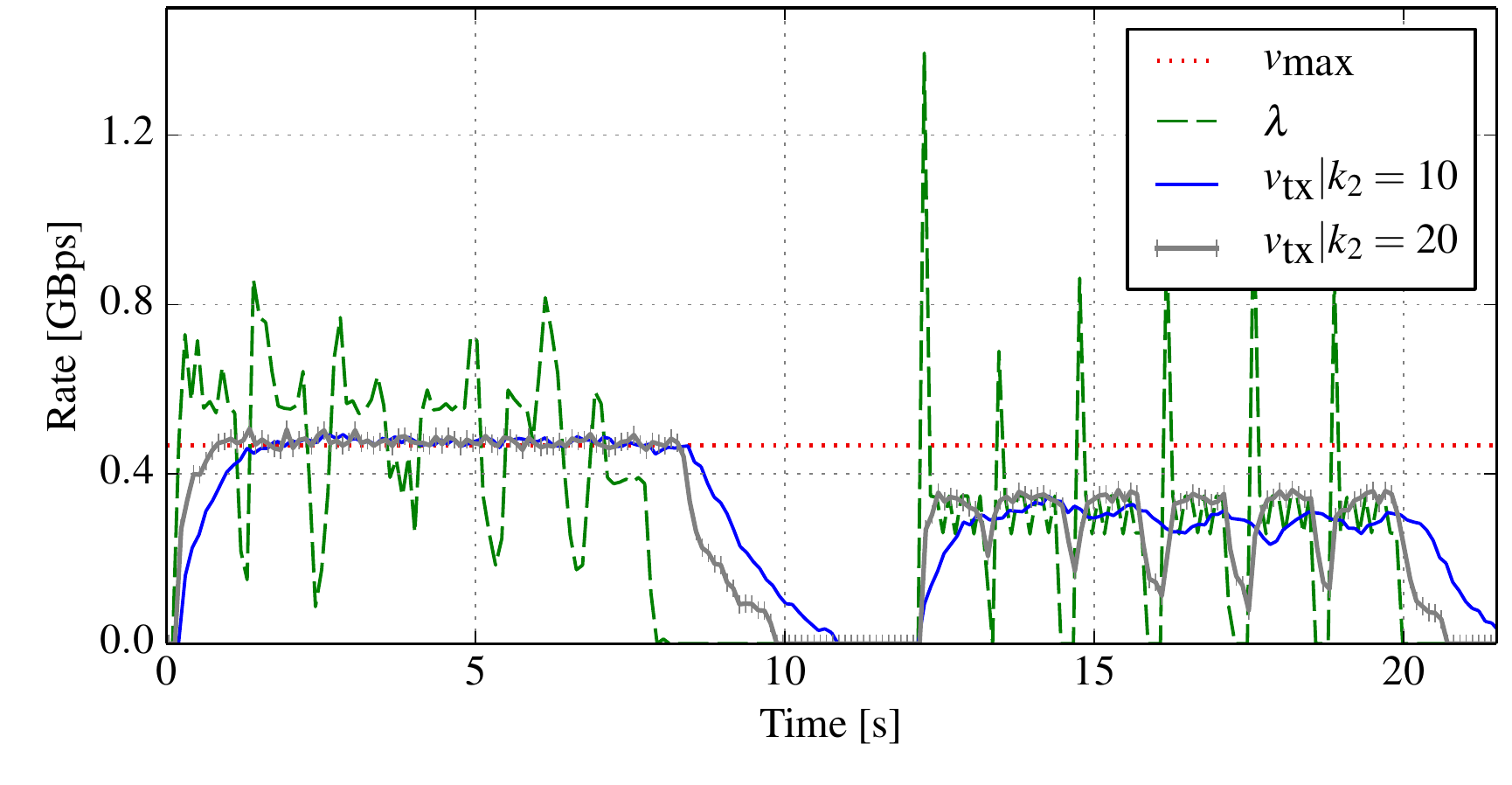}
\caption[Programmable the filtering behaviour]{
{Traffic shaping effects of a CA under two configurations: The same input
traffic pattern $\lambda$ was the input of
a CA configured as \texttt{Rnet1}, first with $k_2=10$~s$^{-1}$ and then with
$k_2=20$~s$^{-1}$. $v_\textrm{tx}$ shows the filtering effects on the
output rate (actual transmissions authorised by the CAs).}
}
\label{fig:expHWEnzyControlCompare}
\end{figure}

For the last part of the experiment, we updated the {last} CA specification,
{re-tuning} its parameters so as to filter even more the traffic bursts (medium
scale frequencies). To do so, we merely modified the values of individual registers 
without re-loading the entire specification (or involving changes in the bitstream).
The {modified} parameters were $k_2=10$~s$^{-1}$ (to reduce
the filtering cut-off frequency), and $e_0=50$~Kmol (to maintain the rate cap
at 0.5~Gbps, since $e_0=v_\textrm{max}/k_2$).
\fig\ref{fig:expHWEnzyControlCompare} shows the difference in the output
behaviour under the same arrival traffic pattern.  The rate capping remained
consistent ($t=[0,8.5~s]$), while the smoothing of burstiness was more
pronounced ($t=[12.5,20~s]$) when $k_2=10$~s$^{-1}$.

In summary this experiment demonstrates both how new algorithms can be
installed in the chemical engine, as well as how a CA can be fine-tuned by
(re-)configuring its parameters. {Both operations are possible at runtime.}


\subsection{Cost-savings from CAs on hardware}\label{sec:expPerformance}
\newtext{
To quantify the advantage of an on-hardware execution of CAs, we looked at the
computational cost involved when executing CAs in the linux kernel.
The rationale behind this measurement is that this computational
overhead/penalty disappears as soon as we move the CAs on the FPGA, and
together with it any delays in packet transmissions due to system load.

We have employed the ChemFlow platform that was used in the experiments of
\cite{jsac13,danms13}, and our metric has been the utilisation of the CPU when
engaged in CA (algorithmic) computations only; omitting related management
tasks (such as servicing of interrupts for the queue management operations, and
monitoring of the chemical engine). 
To understand how the CPU load scales, we measured two CAs of different
complexity, \texttt{Rnet1} and \texttt{Rnet2} (\texttt{Rnet1} has double amount
of species and reactions than \texttt{Rnet2}), and we also varied for each of
them the input rate of events that they processed.

The results for both CAs are plotted in \fig\ref{fig:performance}, grouped by
the rate of input events. We can see that beyond a certain {rate} (1M mol/s) the
cost increases dramatically, eventually stealing the CPU from other
(application) tasks in the system. Thus, even simple CAs are 
computationally expensive! Doubling the amount of occupied chemical resources
does not really double the load, but nevertheless increases it significantly
($\sim10\%$).

\begin{figure}[t]
\centering
\includegraphics[scale= 0.5]{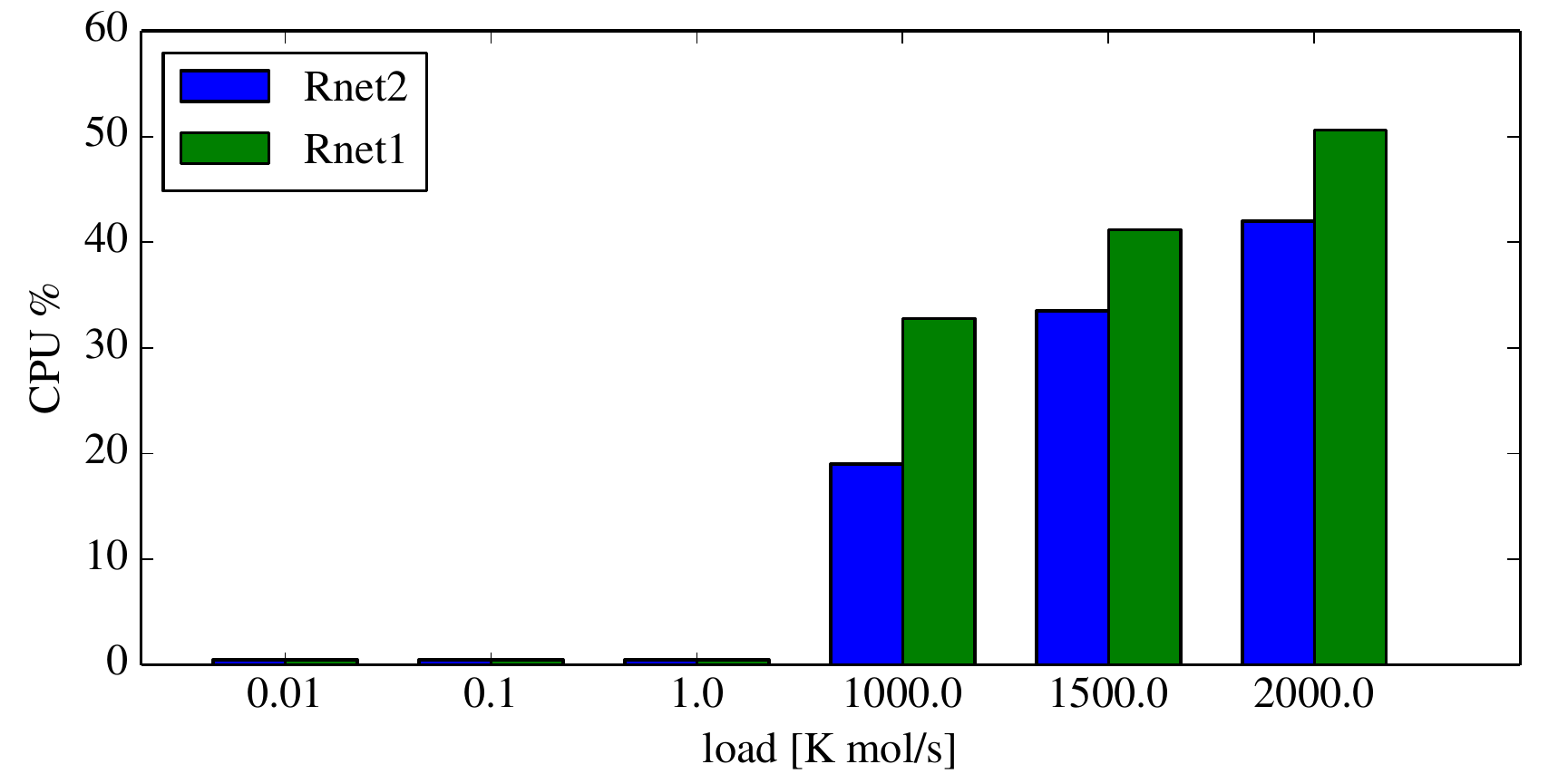}
\caption[Performance evaluation]{
{CPU utilisation when executing directly on the main CPU (as software task in
kernel space of the OS) the CAs in \texttt{Rnet1} and \texttt{Rnet2} with 
different input loads. Monitored for 20 seconds and then averaged separately
for each input load (\newtext{until 1M mol/s load the host CPU utilisation is near zero}).}
}
\label{fig:performance}
\end{figure} 

In the case of the on-hardware implementation, there still exists a saturation
point where the input rate hits the limit of the FPGA clock. In our
implementation, clocking the FPGA at 80~MHz and using 1 mol/KB resolution, this
limit is approximately at 800~Mbps when two reactions are involved in the CA,
and at 1.6~Gbps when there is one reaction involved. Even with such a low-end
FPGA, today these effected speeds are well within norm for edge connectivity,
access networks, and corporate LAN infrastructures (where traffic shaping is
mostly needed). As we climb up the range of higher-end FPGAs and dedicated OS
interfaces (e.g., PCIe bus), there is substantial improvement in performance
(see \sect\ref{sec:FPGAdimensioningHW}) that can serve application needs even
deeper in the core of the network. 
Finally, the scaling of CAs' complexity is not a problem in the case of on-hardware implementation thanks to
parallelisation. The only limitation can be the size of the FPGA (in terms of
number of cells).
}


\section{Discussion}\label{sec:discussion}

\newtext{

To support our initial claim that CAs are well suited to develop a broad range
of control functions for network dynamics, we start this section by providing a
couple of algorithms related to queue scheduling and AQM
(together with others already presented in past
literature~\cite{sensorj14,jsac13,danms13}, they provide a comprehensive
account for network dynamics functions that includes queuing disciplines, AQM,
rate control, distributed access, traffic conditioning, distributed consensus,
and flow control). We then discuss design extensions based on the OpenFlow
architecture~\cite{McPa08}, to illustrate the actual contextualisation in SDN.
We finish the discussion with a reference to tradeoffs and performance
expectations of running the CA framework on various FPGA devices currently
available on market.
}


\subsection{Chemical controllers for Active Queue Management (AQM)}
\label{sec:aqm}

\begin{figure}[t]
   \center
   \begin{minipage}[b]{.52\linewidth}
     \centering
     \centerline{\includegraphics[width =4.5cm]{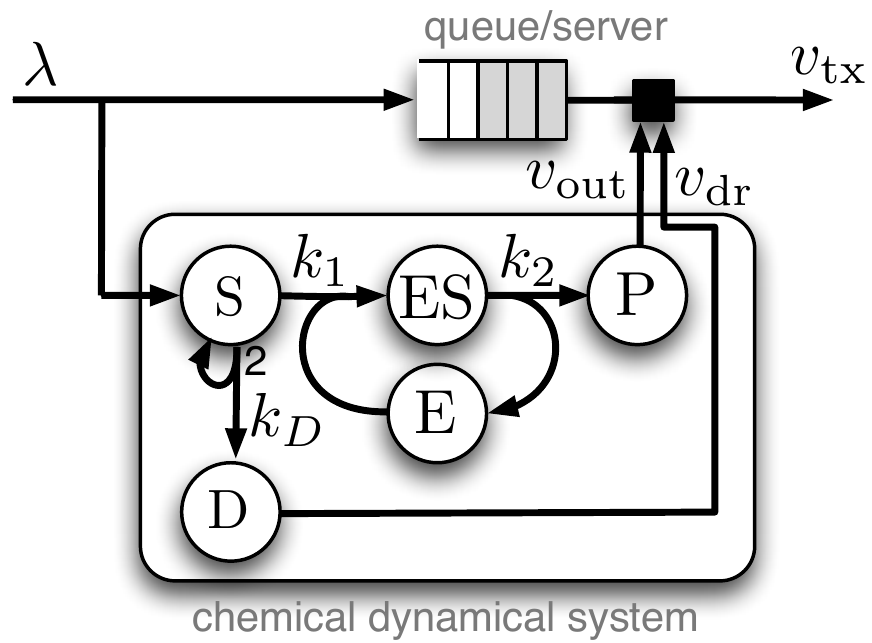}}
     \centerline{(a) System}
   \end{minipage}
   \begin{minipage}[b]{0.45\linewidth}
     \centering
     \begin{subequations}
       \begin{align}
  	 r_{1}:\,\,  & \mol{S}+\mol{E}	 \stackrel{k_1}{\longrightarrow} \mol{ES}	\notag	\\ 
  	 r_{2}:\,\, 	& \mol{ES} 	 \stackrel{k_2}{\longrightarrow}\mol{E} + \mol{P}\notag \\
  	 r_{3}:\,\, 	&2\; \mol{S} 	 \stackrel{k_D}{\longrightarrow}\mol{S} + \mol{D}\notag
       \end{align}
       \vspace{0.3in}
     \end{subequations}
     \centerline{(b) Reaction set}
   \end{minipage}
   \caption[Active Queue Management]{
{\texttt{Rnet3}: The enzymatic reaction network can be extended to be used as a
AQM scheme. The CA has two outputs: species \mol{P} controls the departure
process, and species \mol{D} regulates the drop process of packets from the
queue. The scheme guarantees a maximum transmission rate of packets while
keeping the queue size (and therefore queueing latency) low.}
}
   \label{fig:aqmScheme}
\end{figure}

\begin{figure}[t]
  \centering
  \includegraphics[scale= 0.5]{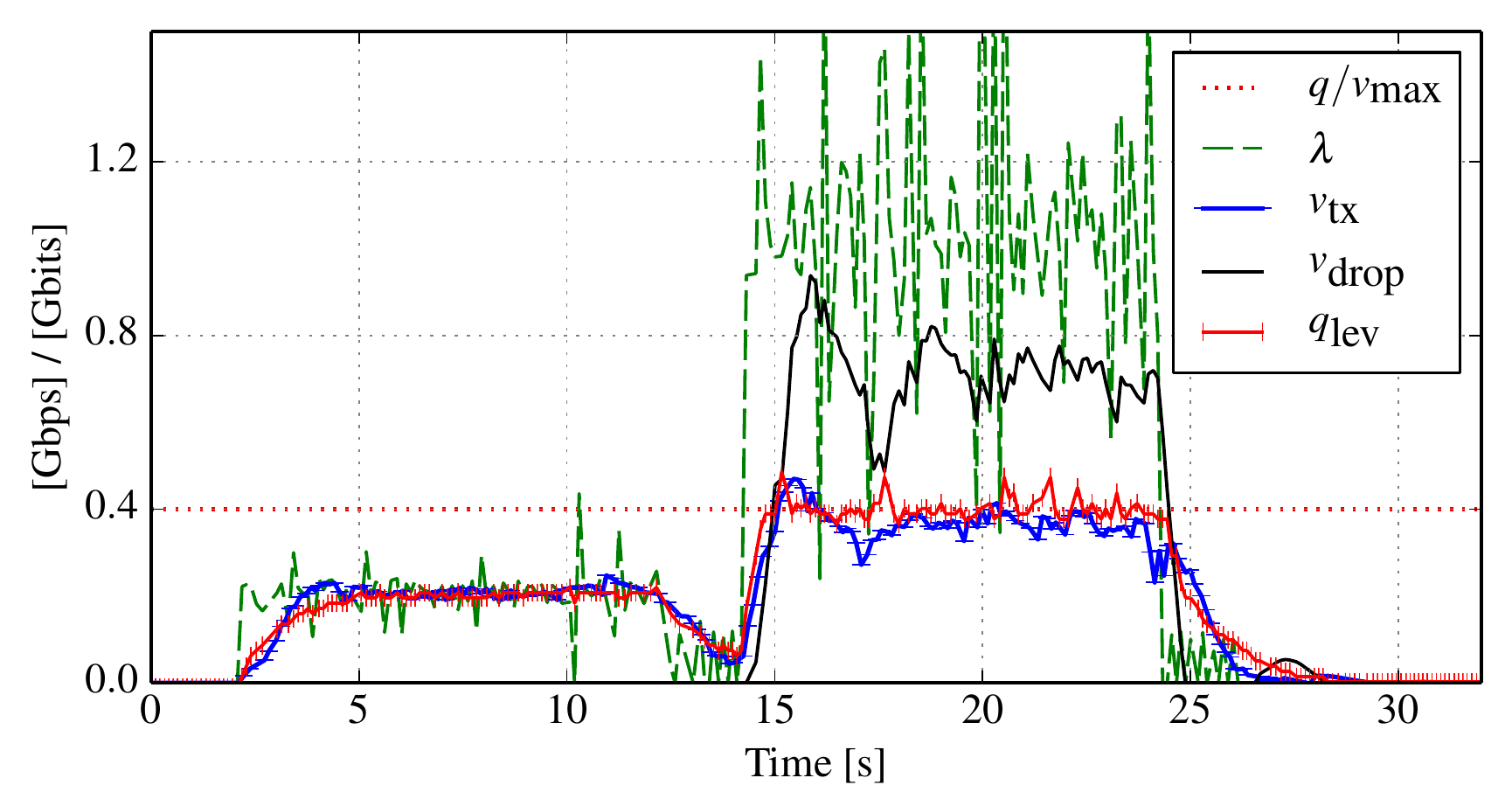}
  \caption[]{
  {Experimental result of an AQM-style chemical controller (\texttt{Rnet3} in \fig\ref{fig:aqmScheme}).}
}
  \label{fig:AQMres}
\end{figure}

\newtext{
A minimal extension of the enzymatic rate controller scheme in
\fig\ref{fig:ChemRateLim} suffices to turn the CA into an AQM scheme with
packet dropping behaviour analogous to RED~\cite{FlJa93}. As shown in
\fig\ref{fig:aqmScheme}, the extension involves one additional reaction
($r_{3}$) and one species (\mol{D}), whose concentration regulates the drop
process at the head of the queue. 

Reaction $r_{3}$ (much slower than $r_{1}$) occasionally ``samples'' the
amount of enqueued packets awaiting transmission (i.e., concentration of
species \mol{S}). If the queue size starts growing (i.e., packets dequeued at too
slow rate or the arrival rate is too high), $r_{3}$ accelerates fast (as a
second order function of the queue size) creating drop tokens (\mol{D}) to
remove packets from the head of the queue. As the queue size decreases, $r_{3}$
quickly recovers again its low speed and effects on queue drops. 

\fig\ref{fig:AQMres} validates experimentally this behaviour in a simple
scenario where \texttt{iperf}-generated VBR UDP traffic goes through a queue
controlled by this CA. The upper rate limit of the enzymatic controller was set
to 0.4~Gbps, representing the maximum desired link utilisation (condition under
which no queue is built-up).  The UDP traffic was admitted to the queue
initially at 0.2 Gbps and then at 1~Gbps, during different phases of the
experiment ($\sim$2s-13s and $\sim$14s-25s).  One can see the drop rate (black
line) being effectively zero under low-load conditions (first phase). As soon
as the rate cap was reached, and the queue started building up (second phase),
the drop-mechanism kicked in emptying the queue at a pace synchronised (no
phase lag) with the queue size variations.

Note that while the CA operates on the queue size (\mol{S} species), its
configuration is in terms of throughput/latency cap (0.4~Gbps) at the queue! In
fact, this is an intuitive/automated configuration approach sought in
modern AQMs~\cite{NiJa12,PaNa12}.
}



\subsection{Chemical controllers for traffic prioritisation}
\label{sec:wfq}

\begin{figure}[t]
\centering
\includegraphics[scale= 0.23]{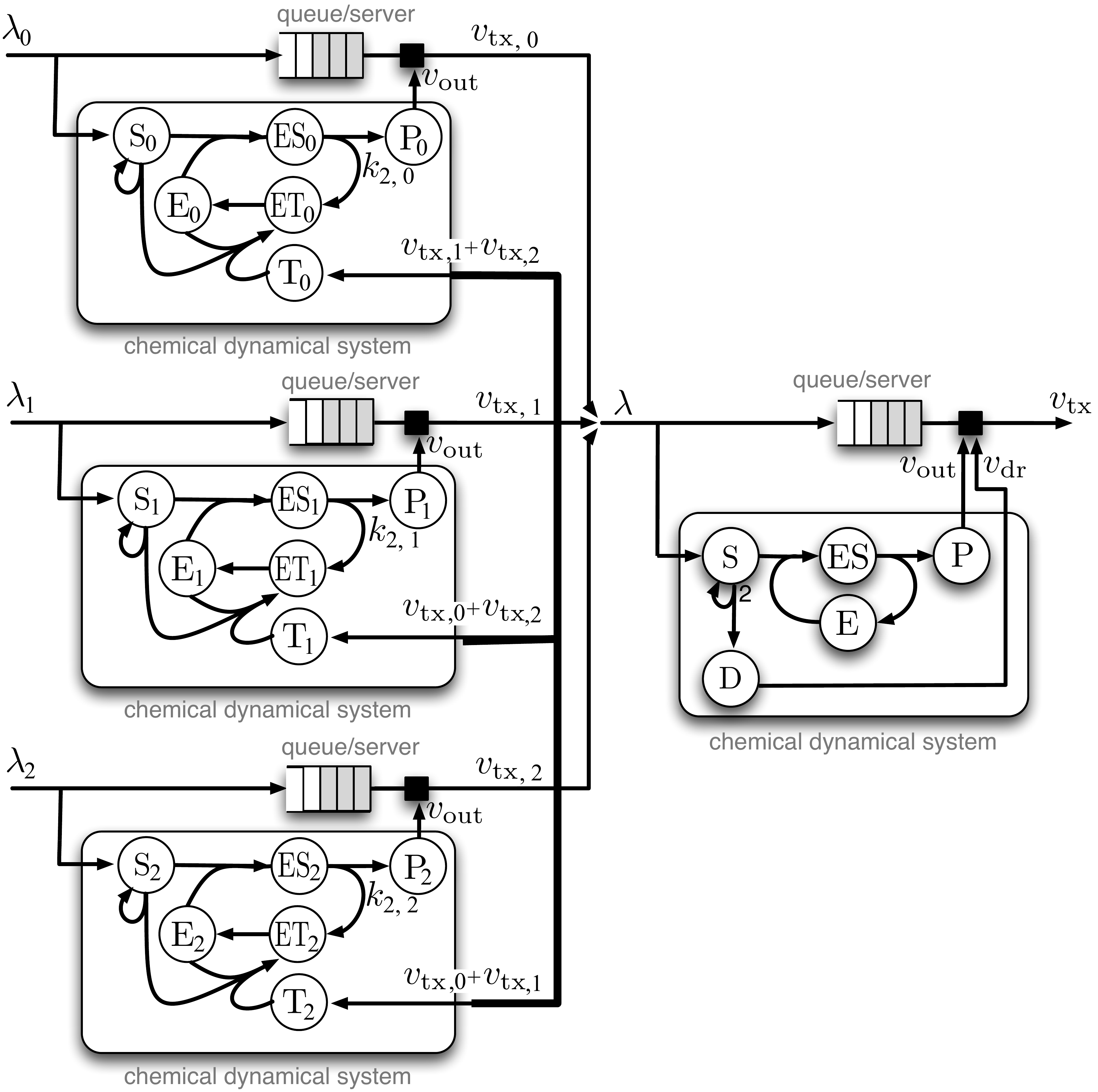}
\caption{
{\texttt{Rnet4}: The combination of \texttt{Rnet3} with the distributed rate
controller scheme in~\cite{danms13} leads to a CA capable of weighted/proportional
fair-queuing. Priorities are configurable via $k_{2,i}$.}
}
\label{fig:prioScheme}
\end{figure}

\begin{figure}[t]
\centering
     \includegraphics[scale=0.5]{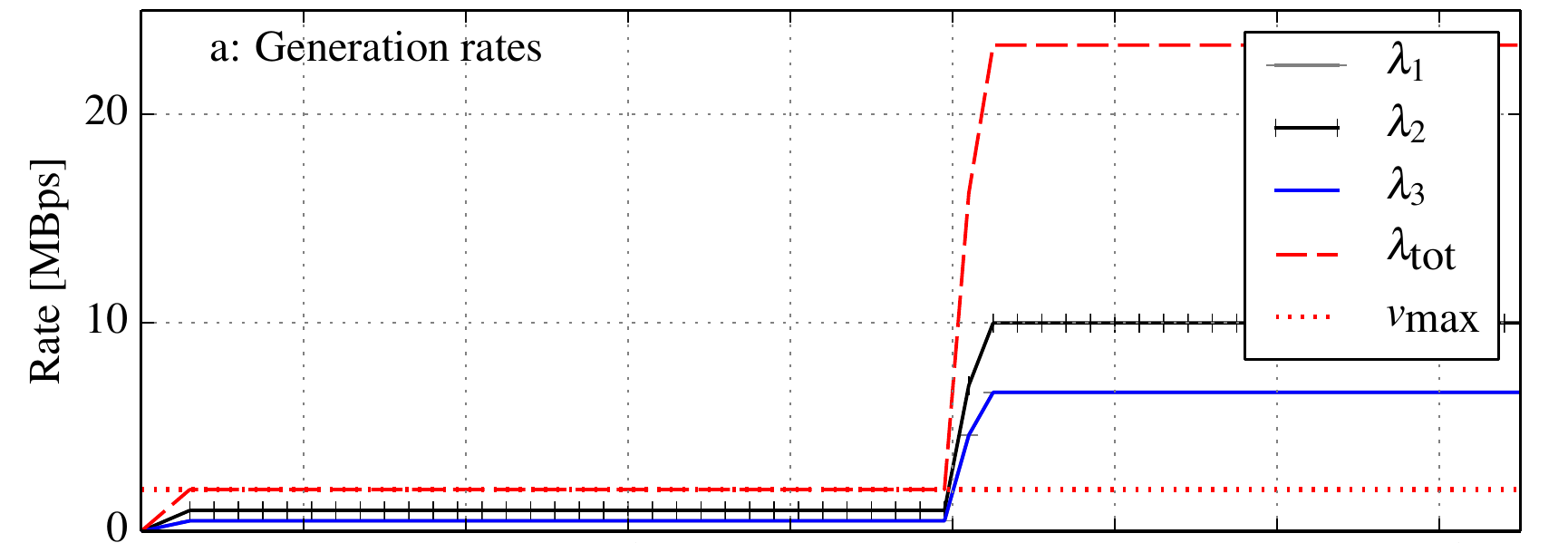}
    \includegraphics[scale=0.5]{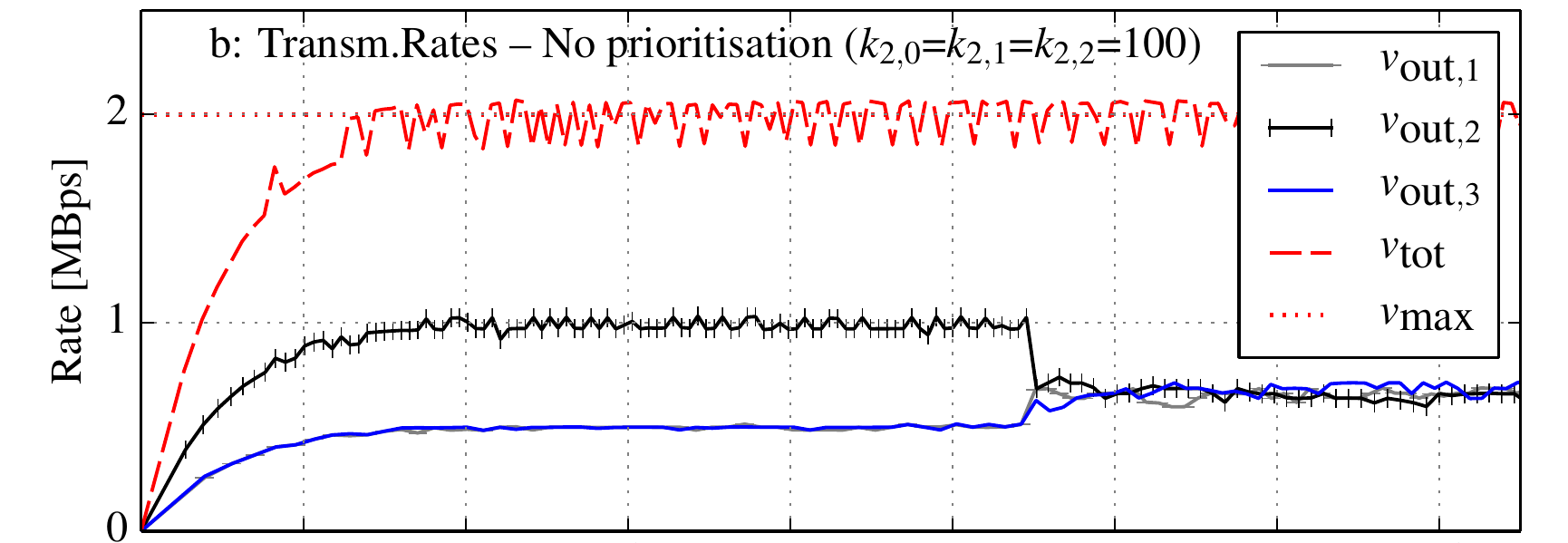}
 	\includegraphics[scale=0.5]{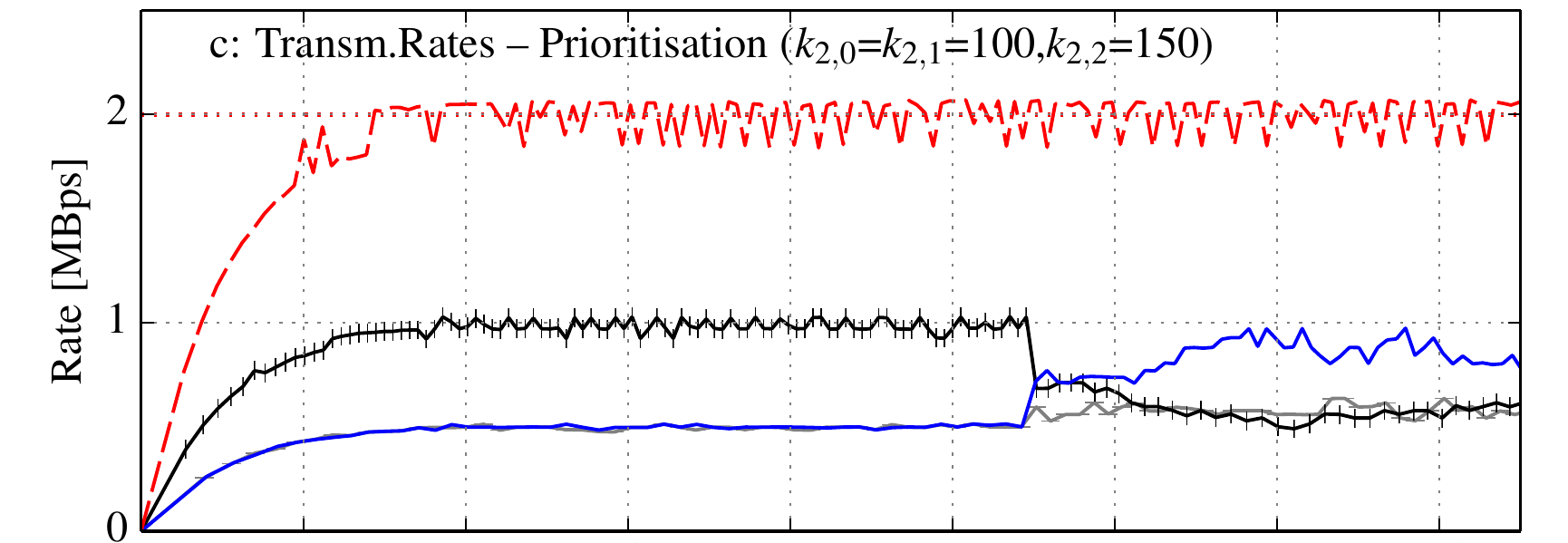}
	     \includegraphics[scale=0.5]{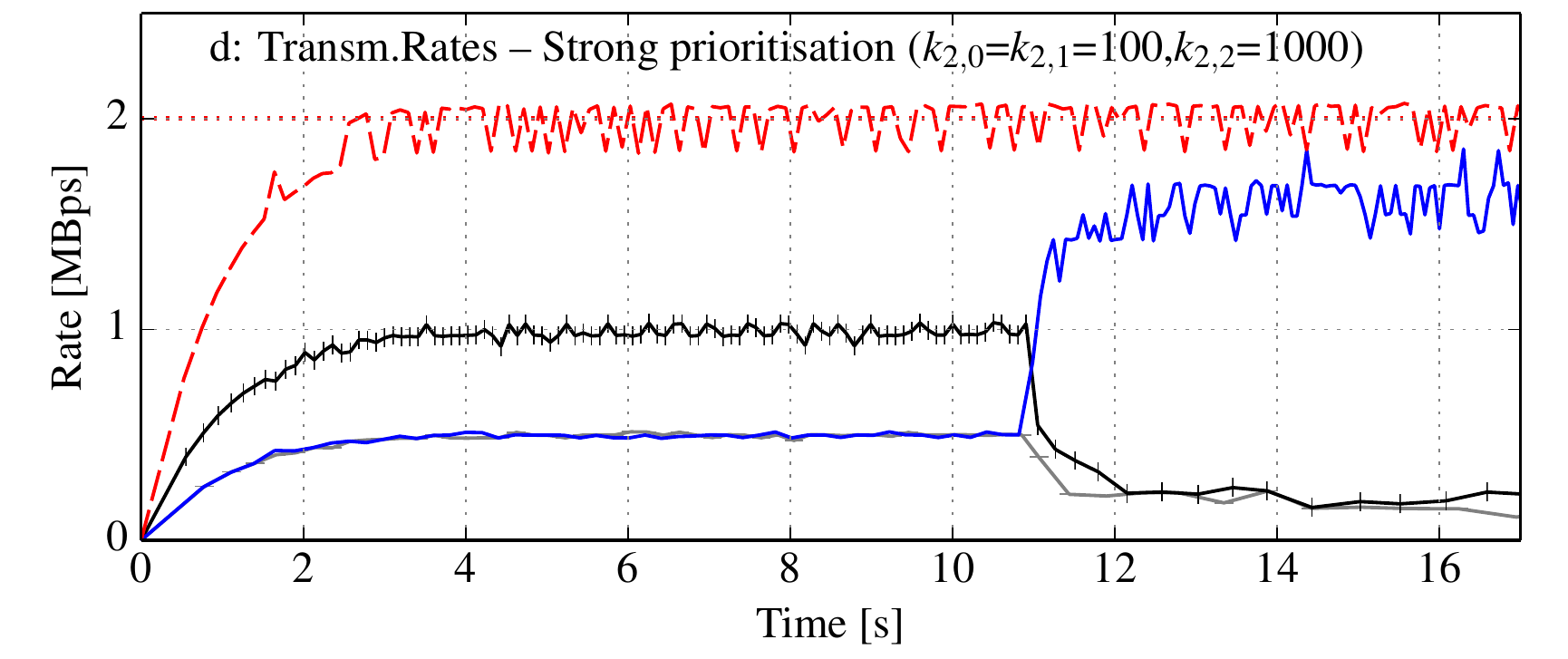}
  \caption{Prioritisation of traffic classes via $k_{2,i}$-values (see CA in
  \fig\ref{fig:prioScheme}). \newtext{Curves $\lambda_1$ and $\lambda_3$ overlap
  and appear as one.} }
  \label{fig:PRIOres}
\end{figure}

\newtext{
By combining the distributed rate control scheme presented in~\cite{danms13}
with the CA for AQM of the previous section, we are able to create a CA for
weighted, or proportional, fair-queuing (\fig\ref{fig:prioScheme}).  The
servers of the participating queues in the scheme (typically corresponding to
distinct classes of traffic) are controlled by identical reaction sub-networks,
sharing their token/molecular state (aggregate of species $\mol{P}_i$ feeds
back to each $\mol{T}_i$). Through coefficients $k_{2,i}$ at each sub-network,
one can configure the proportional bandwidth shares for each queue.  The
outputs of these queues then aggregate at a single egress queue, which is
controlled by the last stage of the CA, a sub-network implementing the AQM
in \sect\ref{sec:aqm}. 

Without delving into analytical details due to space
limitation,\footnote{\newtext{The analysis is a straightforward product of the theory
in~\cite{jsac13}}.} we show an experimental validation in \fig\ref{fig:PRIOres}.
The service processes of three intermediate queues and the egress queue (where
they aggregate) were controlled by the reaction network \texttt{Rnet4}.
\fig\ref{fig:PRIOres}(a) shows the CBR admission rates of traffic, in two phases ($t<10$s
and $t \ge 10$s), to the three queues ($\lambda_1$ and $\lambda_3$
flows had the same rate). \fig\ref{fig:PRIOres}(b/c/d) demonstrate fair-sharing and
weighted (proportional) fair-sharing by means of different $k_{2,i}$
setting.
In the first phase, the total aggregate admission rate (at the intermediate
queues) did not exceed the configured 2Mbps-limit at the egress queue.
All flows claimed and received what they needed from the available
bandwidth.
In the second phase, the total aggregate admission rate exceeded by
far the rate limit and prioritisation kicked in. The share each flow
received is (statistically) proportional to the weights expressed as
$k_{2,i}$ parameters. 
 
It is worth mentioning that CAs of this size become very fast prohibitive for
execution in CPU at the host OS.
}
%


\subsection{CAs and Software Defined Networking}

\begin{figure}[t]
\centering
\includegraphics[height= 5.5cm]{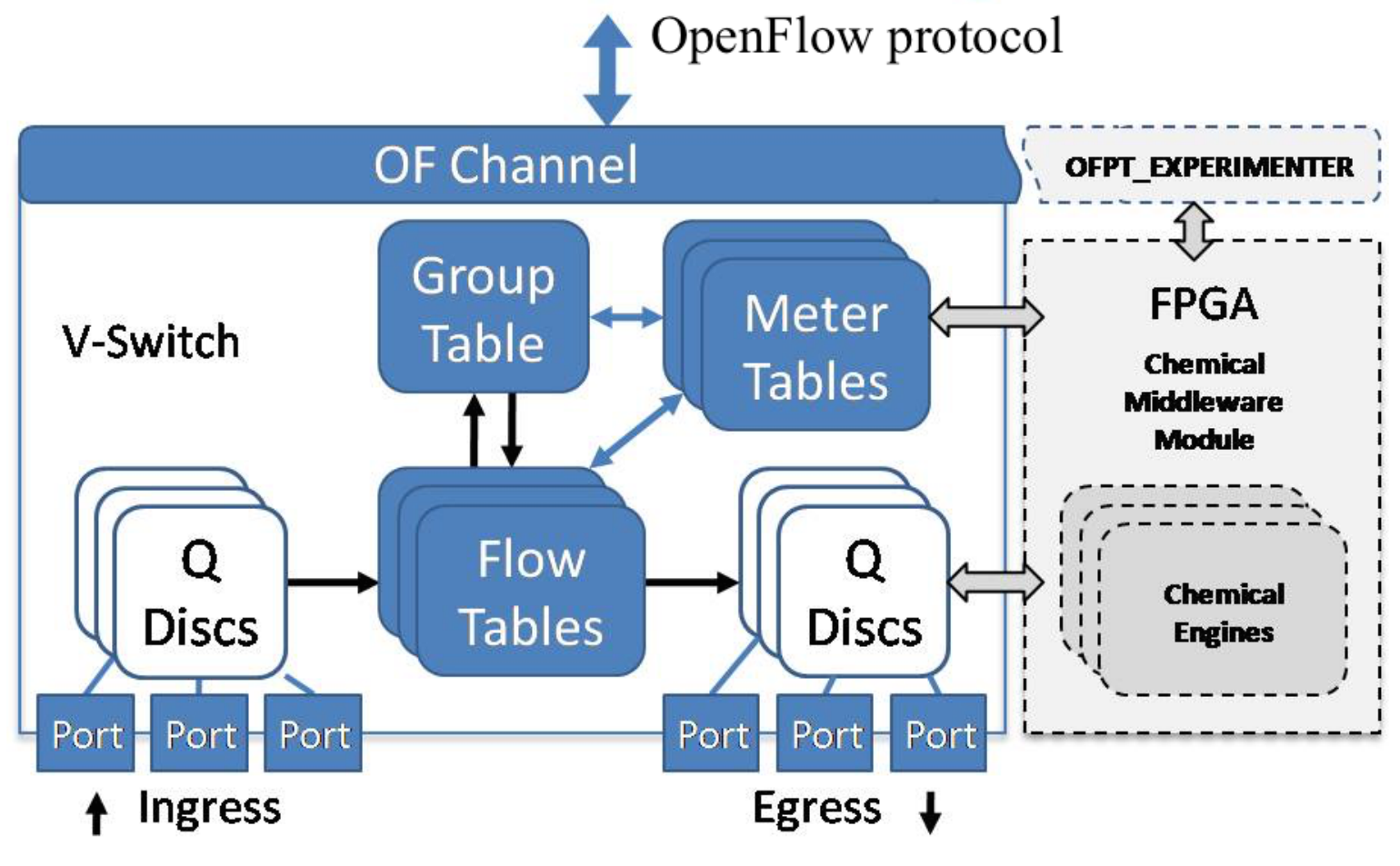}
\caption{
{Integration of CAs in the OpenFlow architecture}
}
\label{fig:sdn}
\end{figure} 

\newtext{ 
In the SDN research landscape deployment of CAs on hardware opens a
door to programmable network dynamics. Looking at the OpenFlow (OF)~\cite{McPa08}
architecture as one of the reference enabling southbound
technologies in SDN,\footnote{\newtext{We are not bound to OpenFlow as a southbound
interface; our choice was driven by its extensibility, broad acceptance by
hardware manufacturers, and its evolution as a melting pot for new
features and capabilities.}}
\fig\ref{fig:sdn} illustrates how the integration of our chemical
framework will be effected in an OF controlled switch.

The dark coloured (in blue) switch subsystems are those currently liable for
remote configuration and re-programming via an OF controller.\footnote{\newtext{This
holds until current version 1.4 of the OF protocol}} One can see that the
queuing subsystem, although available for reading statistics through the OF
protocol, cannot be remotely controlled or modified by an OF controller. Additionally, the only traffic shaping/management feature supported in this architecture
is instantaneous per-flow rate policing. However, the OF architecture accounts
already for more than 40 counters and meter bands (collecting state information and
statistics), which is all a CA requires as inputs.

As shown in \fig\ref{fig:sdn}, an FPGA-based ``chemical subsystem'', like the
one presented in this paper, can be hosted at any OF switch on NetFPGA or other
FPGA-enabled network cards, or on typical manufacturer-provided FPGA boards
wired through hardware interrupts to the OS.  Internally (horizontal interface),
it should be ``permanently'' interfaced with the queuing subsystem; on one side
controlling the enqueue, dequeue (queue server) and head-drop primitives, and
on the other side controlling the increment/decrement primitives of dedicated
I/O species (registers). In a similar fashion, it can also be
``non-permanently'' (programmatically/on-demand) interfaced with the
counter-set of the OF-switch which may be used as additional input species.
These are all mere interrupt signals.

Remote access (southbound SDN interface) from an OF controller is effected
through the OF protocol's experimental extensions. Very simple primitives as
in~\cite{SiSc03} can provide admission control of the chemical engines, as well
as loading and reseting of algorithms by means of reaction network
specifications (or partial specifications providing incremental updates
and modifications for existing reaction networks).  
}


\subsection{Which logic device?}\label{sec:FPGAdimensioningHW}


For our experimentation we have used a low-end FPGA device (XC6SLX9, the 2nd
smallest device of the Spartan-6 family), in which we have exhausted most of
the available logic (70\% of slice LUTs)
%
Yet, we were able to implement a powerful chemical engine that accommodates
up to 255 species and up to 8 reactions, with a maximum of 8 reactant and
product species (sufficient for a number of practical
CAs). This implementation uses a single reaction scheduler (LoMA core) for all
8 reactions, which computes propensities through a linear pipeline of
multiplier DSPs.


\newtext{
By using the XC7K325T FPGA, which is currently mounted on the popular
NetFPGA-1G-CML board, any concern on logic resource exhaustion instantly
vanishes: the same implementation of the chemical middleware framework with an
instantiation of the same amount of chemical resources would utilise barely 1\%
of slice LUTs available.  Indicatively, in Table~\ref{tab3} we provide
summarised reports from the EDA software of Xilinx, for the amount of logic
resources required on the XC7K325T FPGA, when instantiating different amounts
of chemical resources in the chemical middleware (up to 256 species,
$|\mathcal{R}|$ reactions, 8 reactants and products, and a single LoMA core
with a linear pipeline). While the size of the LoMA core is fixed (e.g., 722 Slices,
798 Slice Reg, 2334 LUTs, 79 LUTRAM, and 4 DSP48E1s on the XC7K325T FPGA), 
the logic utilisation scales up as a function of the maximum
amount of chemical resources one is willing to make available in the system for
CAs (e.g., number of reactions, of species, order \etc).
}

\begin{table}[t]
\centering
{\footnotesize
\begin{tabular}{r l l l l}
\hline
           & $|\mathcal{R}|$ = 2 & $|\mathcal{R}|$ = 4 & $|\mathcal{R}|$ = 8  & $|\mathcal{R}|$=32 \\
\hline

\# Slice Registers  &  1'338 & 1'533 & 1'922 & 4'290 \\

\# Slice LUTs           &  3'071& 3'464 & 3'931 & 7'838 \\

\# Occupied Slices    &  1'145 & 1'340 & 1'792 &4'398 \\


\# DSP48E1s          & 4 & 4 & 4 &4\\
\hline
\end{tabular}
}
\caption
{
Logic resource requirements on XC7K325T FPGA for a chemical
framework with up to $|\mathcal{R}|$ reactions available.
}
\label{tab3}
\end{table} 


Next, our system implementation is optimised for economy in logic resources, at the
cost of speed. Operationally, it has been tested at 40 and 80 MHz but it can
also work at 160 MHz, and with optimising differently the circuitry it would be
possible to use at 320 MHz.\footnote{\newtext{The presets of 40, 80, 160 or
320 MHz are the allowed clocking frequencies of the XC6SLX9 FPGA, as specified
by the manufacturer.}} 
At 40 MHz and 80 MHz clock, the system can process external events (e.g. packet
arrivals) that occur every $\sim$ 10 $\mu$s and $\sim$ 5 $\mu$s respectively
(and would be capable of handling $\sim$ 2.5 $\mu$s with 160 MHz clock and down
to $\sim$ 1.2 $\mu$s with a 320 MHz clock). Moreover, it is able to process
correctly two \emph{sporadic} events occurring 50~ns apart, so long as they
last at least $\sim 5$~ns each.
\newtext{
FPGAs with higher clocking frequencies would provide even better resolution,
e.g. the XC7K325T-2 FPGA with up to 650 MHz clocking frequency would allow down 
to 615 ns resolution. Overall, higher clocking frequency means implementing the the LoMA core with faster
DSP modules.
}


Given a certain FPGA, one can further optimise for speed, at the cost of logic
resource economy, by improving drastically on parallelisation. First, one can
employ multiple LoMA cores (up to dedicating one to each reaction). Second,
one can employ in the design of the LoMA core a logarithmic pipeline of DSPs (for
the computation of propensities).
\newtext{
We have experimented with such a design on the XC7K325T-2 FPGA. By dedicating a
LoMA core to each reaction in the configuration of column 4 in Table~\ref{tab3},
the number of clk-cycles for re-scheduling the reactions dropped from
$\sim$1600 to 52, while the logic resource budget increased to 30'055 slice
registers, 79'016 slice LUTs, and 128 DSP48E1s. By additionally changing the
pipeline of the LoMA core, we attained a further reduction to only 24
clk-cycles, and a further increase in logic consumption to 54'154 slice registers,
113'495 slice LUTs, and 320 DSP48E1s. This is still less than 50\% of the logic
resources available on the XC7K325T-2 FPGA, and with a clocking frequency of
400 MHz we have an impressive 60~ns resolution.

At this point however, it is worth noting that these improvements on speed do
not necessarily imply better algorithmic performance. Since the hosted CAs are
dynamical systems there is always a performance trade-off for each CA between
speed of convergence and region of stability, and the choice is application
specific (see \cite{jsac13} for details).
}


\section{Conclusion}\label{sec:Conclusion}

We have introduced, implemented and evaluated a framework that enables run-time
(re-)programmable algorithms on FPGA hardware.  These algorithms, which are
inspired and based on laws and principles of Chemistry, are particularly suited
to functions featuring control of network dynamics. 

\newtext{
The very simple high-level representation of these algorithms (as chemical
reaction networks)

\begin{itemize}
\item has allowed the expression of accurate mathematical models directly on hardware
without the need for low-level HDL programming or even finite state automata,

\item leads to fully parallelisable implementations, where parts of an algorithm can be
modified separately and independently of the rest of the program,

\item has enabled their programmability and configurability on hardware at
sub-second latencies and without the need to field re-program FPGAs.
\end{itemize}

While functions for network dynamics is merely our playground, these algorithms
may describe also user application logic, performing calculations for datasets
other than packets in queues. Hence, in our understanding, this work entails
a promising prospect for on-demand offloading general numerical logic directly on FPGA hardware,
previously only flexibly expressed at the application level and within the
overheads of an operating system. This obsoletes the
need for the less performance- and power-efficient
VonNeumman-architecture-based CPUs and GPUs.
}

\bibliographystyle{IEEEtran}

\end{document}